\documentclass[12pt,english]{article}
\usepackage{times}
\usepackage[T1]{fontenc}
\usepackage{geometry}
\usepackage{rotating}
\usepackage{color}
\usepackage{graphicx}
\usepackage{setspace}
\usepackage{amssymb}

\makeatletter

\providecommand{\tabularnewline}{\\}

\usepackage{bm}
\usepackage{eurosym}

\usepackage{babel}
\makeatother
\begin{document}

\section*{Towards Real-World Indoor Smart Electromagnetic Environments - A
Large-Scale Experimental Demonstration }

\noindent ~

\noindent \vfill

\noindent A. Benoni,$^{\left(1\right)}$ F. Capra,$^{\left(1\right)}$
M. Salucci,$^{\left(1\right)}$ \emph{Senior Member}, \emph{IEEE},
and A. Massa,$^{(1)(2)(3)(4)}$ \emph{Fellow, IEEE}

\noindent \vfill

\noindent {\large ~}{\large \par}

\noindent {\small $^{(1)}$} \emph{\footnotesize ELEDIA Research Center}
{\footnotesize (}\emph{\footnotesize ELEDIA}{\footnotesize @}\emph{\footnotesize UniTN}
{\footnotesize - University of Trento)}{\footnotesize \par}

\noindent {\footnotesize DICAM - Department of Civil, Environmental,
and Mechanical Engineering}{\footnotesize \par}

\noindent {\footnotesize Via Mesiano 77, 38123 Trento - Italy}{\footnotesize \par}

\noindent \textit{\emph{\footnotesize E-mail:}} {\footnotesize \{}\emph{\footnotesize arianna.benoni,
federico.capra, marco.salucci, andrea.massa}{\footnotesize \}@}\emph{\footnotesize unitn.it}{\footnotesize \par}

\noindent {\footnotesize Website:} \emph{\footnotesize www.eledia.org/eledia-unitn}{\footnotesize \par}

\noindent {\footnotesize ~}{\footnotesize \par}

\noindent {\footnotesize $^{(2)}$} \emph{\footnotesize ELEDIA Research
Center} {\footnotesize (}\emph{\footnotesize ELEDIA}{\footnotesize @}\emph{\footnotesize UESTC}
{\footnotesize - UESTC)}{\footnotesize \par}

\noindent {\footnotesize School of Electronic Science and Engineering,
Chengdu 611731 - China}{\footnotesize \par}

\noindent \textit{\emph{\footnotesize E-mail:}} \emph{\footnotesize andrea.massa@uestc.edu.cn}{\footnotesize \par}

\noindent {\footnotesize Website:} \emph{\footnotesize www.eledia.org/eledia}{\footnotesize -}\emph{\footnotesize uestc}{\footnotesize \par}

\noindent {\footnotesize ~}{\footnotesize \par}

\noindent {\footnotesize $^{(3)}$} \emph{\footnotesize ELEDIA Research
Center} {\footnotesize (}\emph{\footnotesize ELEDIA@TSINGHUA} {\footnotesize -
Tsinghua University)}{\footnotesize \par}

\noindent {\footnotesize 30 Shuangqing Rd, 100084 Haidian, Beijing
- China}{\footnotesize \par}

\noindent {\footnotesize E-mail:} \emph{\footnotesize andrea.massa@tsinghua.edu.cn}{\footnotesize \par}

\noindent {\footnotesize Website:} \emph{\footnotesize www.eledia.org/eledia-tsinghua}{\footnotesize \par}

\noindent {\small ~}{\small \par}

\noindent {\small $^{(4)}$} {\footnotesize School of Electrical Engineering}{\footnotesize \par}

\noindent {\footnotesize Tel Aviv University, Tel Aviv 69978 - Israel}{\footnotesize \par}

\noindent \textit{\emph{\footnotesize E-mail:}} \emph{\footnotesize andrea.massa@eng.tau.ac.il}{\footnotesize \par}

\noindent {\footnotesize Website:} \emph{\footnotesize https://engineering.tau.ac.il/}{\footnotesize \par}

\noindent \vfill

\noindent \emph{This work has been submitted to the IEEE for possible
publication. Copyright may be transferred without notice, after which
this version may no longer be accessible.}

\noindent \vfill

\newpage
\section*{Towards Real-World Indoor Smart Electromagnetic Environments - A
Large-Scale Experimental Demonstration }

~

\noindent ~

\noindent ~

\begin{flushleft}A. Benoni, F. Capra, M. Salucci, and A. Massa \end{flushleft}

\vfill

\begin{abstract}
\noindent To the best of the authors' knowledge, this work presents
the first large-scale indoor experimental assessment of an implementation
of the emerging \emph{Smart ElectroMagnetic Environment} (\emph{SEME})
paradigm, which is based on the deployment of static-passive \emph{EM}
skins (\emph{SP-EMS}s) to enhance the coverage in a 5 {[}GHz{]} \emph{Wi-Fi}
network. Unlike standard (laboratory-based) validations reported in
the state-of-the-art (\emph{SoA}) literature, the scenario at hand
mimics a realistic indoor environment to replicate as close as possible
the user experience when using commodity devices. Representative results
from the experimental field trials are reported to confirm the performance
predictions arising from the numerical studies and the tolerance analyses
carried out with a commercial ray-tracing (\emph{RT}) tool. Besides
experimentally validating the \emph{SEME} idea, this study is also
aimed at (roughly) quantifying the economic advantage of a \emph{SEME}
implementation, relying on simple-manufacturing/low-cost field manipulating
devices without any additional biasing circuitry, with respect to
standard approaches that imply the densification of the active radiating
sources.

\noindent \vfill
\end{abstract}
\noindent \textbf{Key words}: Smart ElectroMagnetic Environment (\emph{SEME}),
Indoor, \emph{Wi-Fi}, Static and Passive \emph{EM} Skins (\emph{SP-EMS}s),
Ray Tracing (\emph{RT}), Experimental Assessment.

\newpage
\section{Introduction }

Recently, there has been an exponential growth of research efforts,
from both academy and industry, to develop innovative concepts and
cutting-edge technologies for fulfilling the ever-increasing demand
for connectivity of current \emph{5G} and future \emph{6G} wireless
systems \cite{Hataria 2021}-\cite{Huang 2020}. Indeed, next generations
wireless networks will have to be efficient and smart enough to realize
the vision of a fully connected world where the \emph{Internet-of-Everything}
(or more the \emph{Internet-of-Emotions}) will ensure seamless and
ubiquitous connectivity at an unprecedented scale \cite{Araghi 2022}\cite{Kayraklik 2022}.
Towards this end, ultra-massive multiple-input multiple-output (\emph{UM-MIMO})
\cite{Zheng 2023}, artificial intelligence \cite{Letaief 2022},
orbital angular momentum multiplexing \cite{Affan 2021}, visible-light
communications \cite{Ndjiongue 2021}, quantum computing \cite{Tosi 2023},
and the extension to the terahertz spectrum \cite{Guo 2021}\cite{Dash 2022}
are currently investigated as promising technological recipes for
efficiently and reliably building future wireless systems \cite{Tang 2021}.
Moreover, traditional wireless infrastructures are currently undergoing
a deep revolution driven by the recent introduction of the so-called
\emph{Smart ElectroMagnetic Environment} (\emph{SEME}) paradigm \cite{Yang 2022}-\cite{Massa 2021}.
While the propagation environment has been traditionally perceived
as an impairment to the propagation of \emph{EM} waves, several \emph{SEME}
proofs-of-concept have shown the possibility to control (also doing
programmable) its \emph{EM} fingerprint, thus bringing unprecedented
new opportunities for enhancing the performance of the whole wireless
system not-only acting on the active transmitting/receiving devices
\cite{Yang 2022}-\cite{Flamini 2022}.

\noindent Innovative \emph{EM} wave manipulating devices such as static-passive
electromagnetic skins (\emph{SP-EMS}s) \cite{Salucci 2023}-\cite{Benoni 2022}
or reconfigurable-passive \emph{EMS}s (\emph{RP-EMS}s), also referred
to as Reconfigurable Intelligent Surfaces (\emph{RIS}s) \cite{Araghi 2022}\cite{Kayraklik 2022}\cite{Ndjiongue 2021}\cite{Di Renzo 2020}\cite{Naeem 2022}\cite{Degli-Esposti 2022}-\cite{Stratidakis 2022},
have been introduced as cost-effective alternatives to conventional
recipes (e.g., increasing the transmit power and deploying additional
network infrastructure such as relays \cite{Tang 2022}) for fulfilling
challenging user-experience (\emph{UE}) requirements, while counteracting
undesired phenomena such as non-line-of-sight (\emph{NLOS}), fading,
and shadowing.

\noindent The use of \emph{EMS}s to yield an adequate wireless coverage
and to improve the quality-of-service (\emph{QoS}) has been investigated
in several indoor \cite{Araghi 2022}\cite{Kayraklik 2022}\cite{Lodro 2022}-\cite{Dai 2020}
and outdoor \cite{Salucci 2023}\cite{Benoni 2022}\cite{Oliveri 2022b}\cite{Sang 2023}-\cite{Trichopoulos 2022}
rich-scattering scenarios hosting wireless communication systems working
in either the sub-6GHz \cite{Araghi 2022}\cite{Salucci 2023}\cite{Benoni 2022}
or the millimeter-wave \cite{Flamini 2022}\cite{Rocca 2022}\cite{Tang 2022}
frequency bands. In a nutshell, \emph{EMS}s are artificially-engineered
surfaces \cite{Barbuto 2022}\cite{Liu 2023} consisting of periodic/aperiodic
arrangements of specifically-designed sub-wavelength-sized unit cells
(\emph{UC}s). The micro-scale descriptors of the \emph{EMS}s can be
locally tuned to manipulate the reflection of the impinging \emph{EM}
waves so that, at macro-scale, it is possible to afford anomalous
reflections towards non-Snell angular directions as well as to generate
focused beams \cite{Rocca 2022}\cite{Benoni 2022} or contoured footprints
\cite{Oliveri 2021}\cite{Oliveri 2022} complying with specific application-driven
requirements.

\noindent On the one hand, \emph{RP-EMS}s have drawn particular interest
thanks to the reconfigurability feature enabled by the presence of
tunable components realized with varactor diodes \cite{Araghi 2022},
liquid crystals \cite{Ndjiongue 2021}, graphene \cite{Dash 2022},
or positive-intrinsic-negative (\emph{PIN}) diode switches \cite{Rains 2022}.
However, \emph{RP-EMS}s have also some cons compared to their {}``simpler''
static implementation (i.e., \emph{SP-EMS}s) such as, for instance,
the finite set of discrete states (i.e., $2^{N}$ states with an $N$-bits
design \cite{Pei 2021}) that implies a reduction of the beam-forming
capabilities and, often, the generation of spurious beams (e.g., $1$-bit
structures \cite{Rains 2022}). Certainly, the number of discrete
states of a \emph{UC} can be increased by employing, for instance,
more \emph{PIN} diodes or interfacing the varactor diodes with digital
to analog converters, but all these countermeasures cause a greater
current consumption and they need more expensive/complex control networks
\cite{Rains 2022}. Moreover, the higher architectural complexity
of \emph{RP-EMS}s turns out in a stronger dependence on the manufacturing
tolerances as well as on the non-ideal characteristics of the embedded
tunable loads. This may cause significant deviations from the nominal/simulated
behavior \cite{Pei 2021}.

\noindent Regardless of static or reconfigurable behaviour, the prototyping
of \emph{EMS}-aided wireless communications and real-world field trials
are still scarce and more tests are necessary towards an industrial
perspective \cite{Sang 2023}\cite{Pei 2021}\cite{Trichopoulos 2022}.
Currently, relatively few measurement campaigns have been documented
in the \emph{SoA} literature to prove that the placement of an \emph{EMS}
onto a building wall could be a cost-effective solution to circumvent
coverage blockages \cite{Rains 2022}. 

\noindent To the best of the authors' knowledge, while accurate physics-driven
path loss models, which overcome traditional {}``simplistic'' mathematical
\emph{EM} predictions, have been validated with dedicated experimental
measurements \cite{Tang 2021}\cite{Tang 2022} and the potentialities
of \emph{RIS}-assisted wireless communications have been experimentally
assessed in both outdoor \cite{Sang 2023}-\cite{Pei 2021} and indoor
\cite{Araghi 2022}\cite{Kayraklik 2022}\cite{Lodro 2022}-\cite{Dai 2020}
scenarios, a more exhaustive set of field trials is still needed to
enable the transition from pre-commercial prototypes to an effective
commercial deployment of a \emph{EMS}-based \emph{SEME} \cite{Sang 2023}.

\noindent As a matter of fact, most of the test-beds described in
the literature are still quite far from a real-world every-day-life
context since \emph{ad-hoc} measurement setups are generally considered.
More specifically, the focus is usually on the optimization of a single
\emph{TX-RX} link in artificially-generated \emph{NLOS} conditions
\cite{Araghi 2022}\cite{Kayraklik 2022}\cite{Lodro 2022} where
the \emph{EMS} is mounted on a tripod or equivalent support \cite{Trichopoulos 2022},
while highly-directive and properly-oriented antennas (e.g., horns
\cite{Pei 2021}\cite{Trichopoulos 2022}) are used to illuminate
and/or to collect the reflected \emph{EM} signal. Furthermore, experimental
validations have been generally limited to mono-chromatic signals
\cite{Pei 2021} and there are not documented large-scale experiments
on the exploitation of \emph{SP-EMS}s in real-world \emph{SEME}s.
Indeed, currently available works are exclusively dedicated to the
exploitation of more complex/expensive 1-bit \cite{Sang 2023}\cite{Pei 2021}\cite{Alexandropoulos 2021}\cite{Amri 2021}
or multi-bit \cite{Kayraklik 2022}\cite{Rains 2022}\cite{Dai 2020}
\emph{RP-EMS}s.

\noindent This paper presents the outcomes of a large-scale experimental
demonstration of a real-world indoor \emph{SEME} enabled by \emph{SP-EMS}s.
The main and innovative contributions of this work over the existing
literature consist in (\emph{i}) the first, to the best of the authors'
knowledge, experimental deployment of cost-effective \emph{SP-EMS}s
to improve the indoor coverage in a pre-existing 5 {[}GHz{]} \emph{Wi-Fi}
network, (\emph{ii}) an in-depth modeling of the \emph{EM} propagation
in complex \emph{EMS}-added indoor environments to numerically predict
the dependence of the \emph{EM} coverage on the key descriptors of
the \emph{SP-EMS} (e.g., \emph{EMS} aperture size and post-manufacturing
positioning), (\emph{iii}) a comparative evaluation of the \emph{Total
Cost of Ownership} (\emph{TCO}) \cite{Christensen 2016} for a wireless
network using \emph{SP-EMS}s along with the analysis of the \emph{QoS}
and \emph{UE} indexes to give the readers a complete picture of the
system performance, not only limited to physical-layer parameters.

\noindent The outline of the paper is as follows. Section \ref{sec:Experimental-Test-Bed-Description}
describes the indoor large-scale test-bed by reporting the results
of its numerical and experimental characterization, as well. The implementation
of a \emph{SP-EMS}s-based \emph{SEME} scheme is then detailed, discussed,
and also experimentally-validated in Sect. \ref{sec:SEME-Test-Bed}
where its pros/cons from an economic point of view are also analyzed.
Eventually, some final remarks and conclusions are drawn (Sect. \ref{sec:Conclusions}).

\section{Test-Bed - Reference Scenario \label{sec:Experimental-Test-Bed-Description}}

The experimental test bed is the \emph{Mesiano} building of the Department
of Civil, Environmental, and Mechanical Engineering (\emph{DICAM})
of the University of Trento, in Trento, Italy (Fig. 1). The building
was a former sanatorium around 1920 and it has been successively converted
to the Engineering Faculty in 1984. It is composed by five floors
covering a total area of approximately $13\times10^{3}$ {[}$\mathrm{m}^{2}${]}.
\emph{Wi-Fi} connectivity is provided by means of \emph{Aruba AP-304}
access points (\emph{AP}s) \emph{}operating in the 5 {[}GHz{]} band
with a maximum transmit power of $P_{TX}=23$ {[}dBm{]} \cite{Aruba-Datasheet}.
As a representative example, the following discussion will be focused
on the second floor, which includes five wide hallways (labeled as
{}``\emph{A}'', {}``\emph{B}'', {}``\emph{C}'', {}``\emph{D}'',
and {}``\emph{E}''), alongside several offices, teaching rooms,
and two stairwells and elevators to access the other floors {[}Fig.
2(\emph{a}){]}.

\subsection{\emph{EM} Modeling}

\noindent To have a reliable prediction of the existing \emph{Wi-Fi}
coverage, the commercial \emph{RT}-based simulator Altair WinProp
\cite{WINPROP} has been used to estimate the received power distribution
on the layer at a constant height of $h=1.2$ {[}m{]} that mimics
the typical area where the hand-held smartphone of a standing person
is located (Fig. 3). Towards this end, an accurate \emph{EM} model
of the scenario at hand has been created in the simulation suite starting
from high-fidelity \emph{CAD} drawings that include the exact location
of each structural element (i.e., walls, windows, and doors) as well
as of the actually deployed \emph{AP}s. The \emph{EM} behavior of
the \emph{AP}s has been modeled by importing, in the setup of the
radiating sources of the \emph{RT} model, the full-wave simulated
pattern of the monopole antenna indicated in the manufacturer data-sheet
\cite{Aruba-Datasheet}.

\noindent Figure 2(\emph{a}) shows the \emph{RT}-simulated absolute
map of the received power, $P_{RX}$, at the carrier frequency of
$f=5.64$ {[}GHz{]} (i.e., \emph{IEEE 802.11} channel $\xi=128$)
on a grid of sampling points uniformly spaced by $\delta_{x}=\delta_{y}=0.25$
{[}m{]} where the position $\mathbf{r}_{\Psi}=\left(x_{\Psi},\, y_{\Psi},\, z_{\Psi}\right)$
of each \emph{AP} (i.e., $\Psi_{\gamma}$, $\gamma\in\left\{ A,\, B,\, C,\, D,\, E\right\} $)
serving the area under test is indicated, as well. As it can be inferred,
the most critical regions in terms of \emph{Wi-Fi} coverage are the
two $27$ {[}m{]} long hallways {}``\emph{A}'' and {}``\emph{B}'',
located in the bottom part of the map, having an extension of $\Lambda\left(A\right)=\Lambda\left(B\right)=67.5$
{[}$\mathrm{m}^{2}${]} as well as, albeit to a lesser extent, the
central hallway {}``\emph{C}'' $\Lambda\left(C\right)=60.5$ {[}$\mathrm{m}^{2}${]}
wide. This is not surprising, since the corresponding \emph{AP}s (i.e.,
$\Psi_{A}$, $\Psi_{B}$, and $\Psi_{C}$) are located \emph{beyond}
the extreme left/right/left walls of the related hallways. 

\noindent To give some insights on the experienced \emph{QoS} of a
user, let us analyze more in detail the distribution of the received
power predicted in the hallway {}``\emph{A}'' area {[}Fig. 4(\emph{a}){]}
by noticing that the same outcomes can be drawn for {}``\emph{B}''
since both zones share almost the same $P_{RX}\left(x,y\right)$ distribution
because of their geometric/\emph{EM} symmetry {[}Fig. 2(\emph{a}){]}.

\noindent By setting the threshold value to $P_{th}=-65$ {[}dBm{]}
for a strong \emph{Wi-Fi} connection and suitable data-transfer rates
for services such as video streaming and voice-over-\emph{IP} (\emph{VoIP})
\cite{Benoni 2022}\cite{Araknis-WiFi}, one can obtain the binary
map in Fig. 4(\emph{b}) where the power level turns out to be below
the target one within the region-of-interest (\emph{RoI}) $\Omega_{A}$.
This latter extends on a non-negligible area of $\Lambda_{Ref}\left(\Omega_{A}\right)=23.13$
{[}$\mathrm{m}^{2}${]} ($\rightarrow$ $\Delta\Lambda_{Ref}^{A}\approx34$
\% being $\Delta\Lambda^{\gamma}\triangleq\frac{\Lambda\left(\Omega_{\gamma}\right)}{\Lambda\left(\gamma\right)}$)
and it is located in the right portion of the hallway, that is, in
the opposite direction to the \emph{AP} $\Psi_{A}$, which is installed
\emph{}beyond \emph{}the left wall in $\mathbf{r}_{\Psi_{A}}=\left(2.00,\,2.48,\,2.88\right)$
{[}m{]} (Fig. 4).

\noindent For completeness, the thresholded map of the second floor
is shown in Fig. 5(\emph{a}) where $\Lambda_{Ref}\left(\Omega_{B}\right)=21.88$
{[}$\mathrm{m}^{2}${]} ($\rightarrow$ $\Delta\Lambda_{Ref}^{B}\approx33$
\%) and $\Lambda_{Ref}\left(\Omega_{C}\right)=0.8$ {[}$\mathrm{m}^{2}${]}
($\rightarrow$ $\Delta\Lambda_{Ref}^{B}\approx1.5$ \%), while there
is an optimal coverage in the remaining areas under test (i.e., {}``\emph{D}''
and {}``\emph{E}'' hallways).

\subsection{Experimental Assessment \label{sub:Measurement-Campaign-and}}

Before proceeding to the manufacturing and the deployment of the \emph{EMS}s
to enhance the \emph{Wi-Fi} coverage in the \emph{RoI}s, a measurement
campaign has been carried out to further assess the simulation outcomes.
To faithfully emulate the \emph{UE} in the existing \emph{Wi-Fi} network,
the measurement tests have been performed by avoiding directional
antennas (e.g., horns) connected to high-end instrumentations mounted
on moving supports as in the \emph{SoA} literature (e.g., \cite{Pei 2021}\cite{Trichopoulos 2022})
since it could result in an optimistic over-estimation of the \emph{QoS}
perceived by the actual users. Otherwise, the map of the received
power has been realized by collecting the data with an hand-held Android
smartphone {[}Fig. 6(\emph{b}){]} and setting the measurement locations,
marked on the floor with paper scotch tape {[}Fig. 6(\emph{a}){]},
to the sampling points of the \emph{RT} solver. To filter out outliers
and the measurement noise, the received power measured at each probing
location has been averaged over a time window of $\Delta t=10$ {[}sec{]}.

\noindent Regardless of several (intentional) non-idealities of the
adopted probing setup, the overall measured coverage map of the entire
floor turns out quite close to the simulated one {[}Fig. 2(\emph{b})
vs. Fig. 2(\emph{a}){]}.

\noindent For instance, let us compare the measured $P_{RX}$ map
in the hallway {}``\emph{A}'' {[}Fig. 7(\emph{a}){]} with the simulated
one in Fig. 4(\emph{a}). Besides the good matching of the distribution
of the received power, which decreases with the distance from the
\emph{AP}, it is worth noting that the existence of the low-coverage
\emph{RoI} $\Omega_{A}$ is confirmed in the measured (thresholded)
map {[}Fig. 7(\emph{b}){]} and, even more, that the extension of this
latter (i.e., $\widetilde{\Lambda}_{Ref}\left(\Omega_{A}\right)=20.13$
{[}$\mathrm{m}^{2}${]}) turns out to be close to the \emph{RT}-predicted
one $\Lambda_{Ref}\left(\Omega_{A}\right)$ {[}Fig. 4(\emph{b}){]}.

\noindent For completeness, Figure 5(\emph{b}) shows the measured
version of the thresholded coverage map where one can infer that $\widetilde{\Lambda}_{Ref}\left(\Omega_{B}\right)=23.93$
{[}$\mathrm{m}^{2}${]} {[}i.e., $\widetilde{\Lambda}_{Ref}\left(\Omega_{B}\right)\approx\Lambda_{Ref}\left(\Omega_{B}\right)${]}
and $\widetilde{\Lambda}_{Ref}\left(\Omega_{C}\right)=0.8$ {[}$\mathrm{m}^{2}${]}
{[}i.e., $\widetilde{\Lambda}_{Ref}\left(\Omega_{C}\right)\approx\Lambda_{Ref}\left(\Omega_{C}\right)${]}
as further check of the \emph{RT}-prediction.

\section{\emph{SEME} Test-Bed \label{sec:SEME-Test-Bed}}

Once the presence of low-coverage \emph{RoI}s in the {}``reference
scenario'' has been detected, an \emph{EMS}-based implementation
of the \emph{SEME} paradigm consists in the selection of the most
suitable \emph{EMS}s design (Sect. \ref{sec:SP-EMSs-Layout-and-Design})
by also predicting their effects (Sect. \ref{sub:EM-Modeling SEME})
and estimating possible performance degradations caused by the (almost
unavoidable) tolerances in both the manufactured prototypes and their
final real-world deployment (Sects. \ref{sub:Sensitivity-Analysis SEME}-\ref{sec:SEME-Experimental-Demonstration SEME}).
Towards a commercial deployment, an analysis from the economic viewpoint
(Sect. \ref{sub:TCO}) is also necessary.

\subsection{\emph{SP-EMS} Design \label{sec:SP-EMSs-Layout-and-Design}}

\noindent Figure 8 provides a \emph{3-D} geometrical sketch of the
\emph{SEME} configuration used to improve the \emph{Wi-Fi} coverage
within the hallway {}``\emph{A}''. The position and the extension
of the \emph{SP-EMS} have been chosen to comply with both \emph{EM}
requirements and architectural constraints by considering, as a realistic
strategy, that the existing \emph{Wi-Fi} infrastructure cannot be
modified. More specifically, the following criteria have been adopted:
(\emph{i}) the reflective aperture of the \emph{EMS}, $\mathcal{S}_{A}$,
should harvest as much as possible \emph{EM} energy from the \emph{AP}
$\Psi_{A}$; (\emph{ii}) the \emph{SP-EMS} must be mounted on the
available wall surface without altering the building rules in terms
of security and maintenance.

\noindent By denoting with $\mathbf{r}_{\mathcal{S}}=\left(x_{\mathcal{S}},\, y_{\mathcal{S}},\, z_{\mathcal{S}}\right)$
and $\mathbf{r}_{\Omega}=\left(x_{\Omega},\, y_{\Omega},\, z_{\Omega}=h\right)$
the barycenter of the \emph{SP-EMS} and the desired (anomalous) focusing
point belonging to the spatial region $\Omega$ at height $h$ where
the received power \cite{Balanis 2016}\begin{equation}
P_{RX}\left(\mathbf{r}\right)=\frac{\lambda^{2}G_{RX}}{8\pi\eta}\left|\bm{\mathcal{E}}\left(\mathbf{r}\right)\right|^{2}\label{eq:Received-Power}\end{equation}
is insufficient for guaranteeing a given \emph{QoS} (i.e., $P_{RX}\left(\mathbf{r}\right)<P_{th}$,
$\mathbf{r}\in\Omega$, $P_{th}$ being the desired coverage threshold),
respectively, the \emph{SP-EMS} has been positioned on the wall just
in front of the \emph{AP}, at the location $\mathbf{r}_{\mathcal{S}_{A}}=\left(0.15,\,2.69,\,2.45\right)$
{[}m{]} {[}Fig. 8(\emph{b}){]}, to ensure \emph{LOS} conditions with
both $\mathbf{r}_{\Psi_{A}}$and $\mathbf{r}_{\Omega_{A}}$ ($\mathbf{r}_{\Omega_{A}}=\left(17.5,\,2.33,\,1.20\right)$
{[}m{]} - Fig. 7), but also avoiding to overlap (even partially) the
doors present in the same site.

\noindent The \emph{SP-EMS} has been then designed starting from the
computation of the incidence, $\left(\theta_{i},\,\varphi_{i}\right)$,
and the reflection, $\left(\theta_{r},\,\varphi_{r}\right)$, angular
directions (Fig. 3)\begin{equation}
\left\{ \begin{array}{l}
\left(\theta_{i},\,\varphi_{i}\right)=\mathcal{F}\left\{ \mathbf{r}_{\Psi},\,\mathbf{r}_{\mathcal{S}}\right\} \\
\left(\theta_{r},\,\varphi_{r}\right)=\mathcal{F}\left\{ \mathbf{r}_{\Omega},\,\mathbf{r}_{\mathcal{S}}\right\} \end{array}\right.,\label{Incidence-Reflection-Angles}\end{equation}
 $\mathcal{F}\left\{ \,.\,\right\} $ being the Cartesian-to-spherical
operator mapping either $\mathbf{r}_{\Psi}$ or $\mathbf{r}_{\Omega}$
to the corresponding elevation and azimuth angles in the local coordinate
system $\mathbf{r}'=\left(x',\, y',\, z'\right)$ centered in $\mathbf{r}_{\mathcal{S}}$
(Fig. 9) \cite{Benoni 2022}. In (\ref{eq:Received-Power}), $\lambda$
($\lambda\triangleq\frac{c_{0}}{f}$) and $\eta$ ($\eta\triangleq\sqrt{\frac{\mu_{0}}{\varepsilon_{0}}}$)
are the wavelength at the working frequency $f$ and the impedance
of the free-space with permittivity $\varepsilon_{0}$ and permeability
$\mu_{0}$, respectively, $c_{0}$ ($c_{0}\triangleq\frac{1}{\sqrt{\varepsilon_{0}\mu_{0}}}$)
being the speed of light, while $G_{RX}$ is the maximum gain of the
receiver and $\bm{\mathcal{E}}\left(\mathbf{r}\right)$ is the electric
field.

\noindent According to (\ref{Incidence-Reflection-Angles}), the values
of the incidence and reflection angles turn out to be equal to $\left(\theta_{i}^{A},\,\varphi_{i}^{A}\right)=\left(14.5,\,116.0\right)$
{[}deg{]} and $\left(\theta_{r}^{A},\,\varphi_{r}^{A}\right)=\left(4.29,\,-106.1\right)$
{[}deg{]}, by also numerically confirming the need for a double-anomalous
(i.e., non-Snell in both elevation and azimuth) reflection to reach
the \emph{RoI}.

\noindent Figure 9 shows the reference monolithic layout adopted for
synthesizing the \emph{SP-EMS}s of the \emph{SEME} test-bed. The \emph{EMS}
aperture $\mathcal{S}$ of side $L$ consists of a set of $N\times N$
metal-backed \emph{UC}s arranged in a uniform square lattice with
half-wavelength inter-element spacing $\ell_{UC}=\frac{\lambda}{2}$
{[}$L=\left(N\times\ell_{UC}\right)${]}. More specifically, the dual-layer
stacked structure in \cite{Huang 2008} has been chosen as the \emph{EMS}
atom to excite a multi-resonant behavior and to yield a smooth phase
variation in a sufficiently wide angular range by using relatively
inexpensive dielectric materials and off-the-shelf thicknesses. To
further keep the design and the manufacturing complexity as low as
possible, the bottom layer and the top one of each \emph{UC} have
been realized by using identical \emph{FR-4} substrates of relative
permittivity $\varepsilon_{r}=4.4$, loss tangent $\tan\delta=0.02$,
and thickness $t=1.6$ {[}mm{]} {[}Fig. 9(\emph{a}){]}, while two
square metallic patches of side $\ell_{1}$ and $\ell_{2}=\left(\alpha\times\ell_{1}\right)$,
$\alpha=0.8$ being a scaling factor \cite{Huang 2008}, have been
etched on the top face of each printed circuit board (\emph{PCB})
layer to enable a local fine-tuning of the reflection phase.

\noindent The skin size $L$ has been chosen as the best trade-off
between the capability of collecting/re-radiating a sufficient amount
of \emph{EM} energy and the overall cost as well as its physical encumbrance,
which should be kept as low as possible to minimize both the economic
and the architectural impact of the \emph{SEME} solution. Towards
this end, the relationship between the \emph{EMS} dimension and the
achievable coverage enhancement has been analyzed. 

\noindent With reference to the hallway {}``\emph{A}'', as a representative
example, $L_{A}$ has been varied within the range $L_{A}\in\left[0.28,\,0.80\right]$
{[}m{]} (i.e., $N_{A}\in\left[10,\,30\right]$) and for each \emph{EMS}
aperture (i.e., the number of \emph{UC}s per side, $N_{A}$), the
two-step synthesis method detailed in \cite{Oliveri 2021} has been
applied to derive the set of $N\times N$ geometric degrees-of-freedom,
$\underline{\chi}=\left\{ \ell_{1}^{mn};\, m,\, n=1,...,N\right\} $,
that model the metallic pattern of the corresponding \emph{EMS} layout.
Figure 10(\emph{a}) summarizes the outcomes of such an analysis by
showing the behavior of the cumulative density function (\emph{CDF})
of the received power inside the hallway at hand\begin{equation}
\Theta\left\{ \left.P_{RX}\left(\mathbf{r}\right)\right|\widehat{P}\right\} =\textnormal{Pr}\left\{ P_{RX}\left(\mathbf{r}\right)\leq\widehat{P}\right\} \label{eq:CDF}\end{equation}
where $\textnormal{Pr}\left\{ \,.\,\right\} $ denotes the probability
function and $\widehat{P}\in\left[-75,\,-40\right]$ {[}dBm{]}. As
it can be inferred, there is always a non-negligible improvement of
the wireless coverage with respect to the {}``reference'' scenario
and, as expected, wider \emph{EMS} apertures correspond to more significant
enhancements. Quantitatively, the probability of being below the threshold
($\Theta_{th}=\Theta\left\{ \left.P_{RX}\left(\mathbf{r}\right)\right|P_{th}\right\} $)
progressively decreases from $\Theta_{th}^{Ref}=45$ \% down to $\Theta_{th}^{SEME}\left(A\right)=12$
\% for the widest \emph{EMS} aperture (i.e., $L_{A}=0.8\,[\mathrm{m}]$),
$\Theta_{th}^{SEME}\left(A\right)=35$ \% being the probability for
the smallest skin ($L_{A}=0.28$ {[}m{]} $\to$ $N_{A}=10$).

\noindent A similar trend holds true for the reduction of the \emph{RoI}
area, $\rho_{SEME}$, defined as\begin{equation}
\rho_{SEME}\left(\Omega\right)\triangleq\frac{\Lambda_{Ref}\left(\Omega\right)-\Lambda_{SEME}\left(\left.\Omega\right|L\right)}{\Lambda_{Ref}\left(\Omega\right)},\label{eq:RHO}\end{equation}
whose behavior versus the \emph{EMS} size $L$ and the corresponding
values are reported in Fig. 10(\emph{b}) and Tab. I, respectively.
It turns out that $\rho_{SEME}\left(\Omega_{A}\right)=37.8\%$ and
$\rho_{SEME}\left(\Omega_{A}\right)=79.5\%$ for the smallest (i.e.,
$L_{A}=0.28\,[\mathrm{m}]$) and the largest (i.e., $L_{A}=0.80\,[\mathrm{m}]$)
\emph{SP-EMS}, respectively.

\noindent By taking into account both \emph{EM} coverage requirements
and installation constraints, the intermediate-size \emph{EMS} with
$L_{A}=0.55$ {[}m{]} ($N_{A}=20$) {[}Fig. 9(\emph{a}){]} has been
selected as the best compromise, since $\Theta_{th}^{SEME}\left(A\right)=18$
\% {[}Fig. 10(\emph{a}){]} and $\rho_{SEME}\left(\Omega_{A}\right)=70\%$
{[}Fig. 10(\emph{b}){]} while wider apertures provide only marginal
improvements.

\noindent For comparison purposes, the coverage enhancement yielded
by installing another \emph{AP} {[}referred as standard (\emph{STD})
approach{]} on the opposite side of the hallway ($\mathbf{r}_{\Psi_{A}^{\left(2\right)}}=\left(29.2,\,2.00,\,2.88\right)$
{[}m{]}) has been predicted, as well. As expected, the use of an active
device allows one to outperform the fully-passive \emph{SEME} solution
in terms of coverage enhancement (i.e., $\Theta_{th}^{STD}\left(A\right)=3$
\% {[}Fig. 10(\emph{a}){]} and $\rho_{STD}\left(\Omega_{A}\right)=95$
\% {[}Fig. 10(\emph{b}){]}), but, besides the need of more transmitting
power and higher economic costs (see Sect. \ref{sub:TCO}), adding
one \emph{AP} may not be a feasible solution in many practical situations
due to the additional cabling and network infrastructure.

\subsection{\emph{EM} Modeling\label{sub:EM-Modeling SEME}}

\noindent A pictorial \emph{3D} representation of the \emph{EM} behavior
of the \emph{SP-EMS} in Fig. 9(\emph{a}) is shown in Fig 11 where
the full-wave simulated \cite{HFSS 2021} directivity pattern is reported.
As it can be observed, the synthesized \emph{EMS} correctly radiates
a pencil beam towards the desired anomalous reflection direction,
the (simulated) maximum directivity being equal to $D_{\max}=30.73$
{[}dB{]}, whose effect is a noticeable increase of the received power
within the entire hallway {}``\emph{A}'' {[}Fig. 12(\emph{a}) vs.
Fig. 4(\emph{a}){]} as pointed out by the difference map $\Delta P_{RX}$,\begin{equation}
\Delta P_{RX}\left(\mathbf{r}\right)\triangleq P_{RX}^{SEME}\left(\mathbf{r}\right)-P_{RX}^{Ref}\left(\mathbf{r}\right),\label{eq:Delta-PRX}\end{equation}
in Fig. 12(\emph{c}). Quantitatively, it turns out that the average
{[}$\Delta P_{RX}^{\mathrm{avg}}\left(A\right)\triangleq\frac{1}{\Lambda\left(A\right)}\int_{A}\Delta P_{RX}\left(\mathbf{r}\right)d\mathbf{r}${]}
and the maximum {[}$\Delta P_{RX}^{\max}\left(A\right)\triangleq\max_{\mathbf{r}\in A}\left\{ \Delta P_{RX}\left(\mathbf{r}\right)\right\} ${]}
power increment are equal to $\Delta P_{RX}^{\mathrm{avg}}\left(A\right)=2.64$
{[}dB{]} and $\Delta P_{RX}^{\max}\left(A\right)=9.60$ {[}dB{]},
respectively (Tab. I).

\noindent From a \emph{UE} perspective, the effect of the \emph{SP-EMS}
is pointed out by the thresholded map in Fig. 12(\emph{b}) where the
\emph{RoI} extension reduces down to $\Lambda_{SEME}\left(\Omega_{A}\right)=6.94$
{[}$\mathrm{m}^{2}${]} ($\rightarrow$ $\Delta\Lambda_{SEME}^{A}\approx10$
\%) from $\Lambda_{Ref}\left(\Omega_{A}\right)=23.13$ {[}$\mathrm{m}^{2}${]}
in Fig. 4(\emph{b}) {[}i.e., $\rho_{SEME}\left(\Omega_{A}\right)=70$
\%{]}.

\noindent For completeness, the results of the \emph{EMS}-based \emph{SEME}
implementations in the zones {}``\emph{B}'' and {}``\emph{C}''
are summarized in Fig. 13 and Fig. 14, respectively, where the corresponding
difference maps and \emph{CDF}s are reported.

\noindent As expected, the distribution of the $\Delta P_{RX}$ values
in the hallway {}``\emph{B}'' is almost symmetrical to that in the
hallway {}``\emph{A}'' {[}Fig. 13(\emph{a}) vs. Fig. 12(\emph{c}){]},
while the related statistical indexes are very close being $\Delta P_{RX}^{\mathrm{avg}}\left(B\right)=2.52$
{[}dB{]} and $\Delta P_{RX}^{\max}\left(B\right)=8.21$ {[}dB{]} (Tab.
I vs. Tab. II). Similar outcomes can be drawn from the behavior of
the \emph{CDF} in Fig. 13(\emph{b}).

\noindent Concerning the {}``\emph{C}'' hallway, the deployment
of the smaller ($L_{C}=0.32$ {[}m{]}) \emph{SP-EMS} in Fig. 9(\emph{b})
as in Fig. 15 {[}$\left(\theta_{i}^{C},\,\varphi_{i}^{C}\right)=\left(15.33,\,28.07\right)$
{[}deg{]} and $\left(\theta_{r}^{C},\,\varphi_{r}^{C}\right)=\left(4.19,\,-43.60\right)$
{[}deg{]}{]} has been enough to fulfill the \emph{QoS} requirements
since $\rho_{SEME}\left(\Omega_{A}\right)=100$ \% being $\Theta_{th}^{SEME}\left(C\right)=0.0$
\% {[}Fig. 14(\emph{b}){]}. Indeed, a lower and limited coverage improvement
than that for the scenario {}``\emph{A}'' ({}``\emph{B}'') {[}Fig.
14(\emph{a}) vs. Fig. 12(\emph{c}) and Fig. 13(\emph{a}){]} was required
to reach the target threshold of $P_{th}=-65$ {[}dBm{]} (e.g., $\Delta P_{RX}^{\mathrm{avg}}\left(C\right)=0.84$
{[}dB{]} vs. $\Delta P_{RX}^{\mathrm{avg}}\left(A\right)=2.64$ {[}dB{]}
and $\Delta P_{RX}^{\mathrm{avg}}\left(B\right)=2.52$ {[}dB{]} -
Tab. I vs. Tab. II).

\subsection{Sensitivity Analysis\label{sub:Sensitivity-Analysis SEME}}

\noindent Before proceeding to the manufacturing and the installation
of the \emph{SP-EMS}s, a sensitivity analysis has been carried out
to predict possible performance degradations caused by the (almost
unavoidable) tolerances in their final deployment in the real test-bed.
Towards this end, a set of positioning inaccuracies has been modeled
by considering an error along the horizontal ($\Delta y$) and/or
the vertical ($\Delta z$) axes {[}Fig. 16(\emph{a}){]}. 

\noindent With reference to the deployment in the hallway {}``\emph{A}'',
Figure 16(\emph{b}) plots the values of the \emph{RoI} reduction,
$\rho_{SEME}\left(\Omega_{A}\right)$, versus the offsets in the range
$\left(\Delta y,\,\Delta z\right)\in\left[-0.1,\,0.1\right]$ {[}m{]}.
Despite the non-negligible change of the incidence/reflection angles
that would be necessary for a re-design of the skin when moving its
barycenter (e.g., $\left(\Delta\theta_{i}^{A},\,\Delta\varphi_{i}^{A}\right)=\left(-0.75,\,17.18\right)$
{[}deg{]} and $\left(\Delta\theta_{r}^{A},\,\Delta\varphi_{r}^{A}\right)=\left(0.41,\,-2.72\right)$
{[}deg{]} when $\Delta y=\Delta z=0.1$ {[}m{]}), the {}``nominal''
\emph{SP-EMS} proves to be robust to the positioning errors. As a
matter of fact, $\rho_{SEME}\left(\Omega_{A}\right)\geq26.9$ \% whatever
the combination of $\Delta y$ and $\Delta z$ values {[}Fig. 16(\emph{b}){]}.

\noindent These conclusions are further confirmed by the plots of
the \emph{CDF} in the presence of either an horizontal {[}$\Delta z=0$
{[}m{]}, $\Delta y\in\left[-0.1,\,0.1\right]$ {[}m{]} - Fig. 16(\emph{c}){]}
or a vertical {[}$\Delta y=0$ {[}m{]}, $\Delta z\in\left[-0.1,\,0.1\right]$
{[}m{]} - Fig. 16(\emph{d}){]} shift of the nominal \emph{EMS} barycenter
since always $\Theta_{th}^{SEME}\left(A\right)$ $\le$ $\Theta_{th}^{Ref}$
and the maximum deviation from the nominal value is around than $15$
\% {[}Figs. 16(\emph{c})-16(\emph{d}){]}, the main impact being due
to the horizontal errors {[}Fig. 16(\emph{d}){]}.

\subsection{Experimental Assessment \label{sec:SEME-Experimental-Demonstration SEME}}

\noindent A prototype of the \emph{SP-EMS} $\mathcal{S}_{A}$ of side
$L_{A}=0.55$ {[}m{]} has been fabricated via \emph{PCB} manufacturing
using off-the-shelf \emph{FR-4} substrates {[}Fig. 17(\emph{a}){]}
and, successively, it has been installed in the experimental test-bed
as modeled in the numerical analyses {[}Fig. 17(\emph{b}) vs. Fig.
8{]}. Three small removable mechanical supports, made of plywood,
fixed to the wall with steel nails (Fig. 17) have been used to easily
mount/remove the \emph{SEME} device for performing several field trials.%
\footnote{\noindent From a practical point of view, it is worth mentioning that
the prototype is quite lightweight (the total weight being approximately
$0.5$ {[}Kg{]}). Therefore, it is easy to deploy a permanent installation
using (for instance) proper glues or specific double sided tapes.%
}

\noindent After the \emph{EMS} deployment, the \emph{Wi-Fi} coverage
(measured) in the hallway {}``\emph{A}'' is shown in Fig. 18(\emph{a}),
while the difference map with respect to the (measured) reference
distribution of Fig. 7(\emph{a}) is reported in Fig. 18(\emph{b}).
Such experimental results confirm the outcomes of the numerical assessment
since, besides the visible improvement of the received power distribution
inside the area under test {[}Fig. 18(\emph{a}) vs. Fig. 7(\emph{a})
and Fig. 18(\emph{b}){]}, there is a close agreement also quantitatively.
As a matter of fact, the measured value of the average increment of
the received power in the complete hallway turns out to be just $0.24$
{[}dB{]} different from the predicted one (i.e., $\Delta\widetilde{P}_{RX}^{\mathrm{avg}}\left(A\right)=2.4$
{[}dB{]} vs. $\Delta P_{RX}^{\mathrm{avg}}\left(A\right)=2.64$ {[}dB{]}).
Similarly, the maximum measured value has been equal to $\Delta\widetilde{P}_{RX}^{\max}\left(A\right)=9.0$
{[}dB{]}, $\Delta P_{RX}^{\max}\left(A\right)=9.60$ {[}dB{]} being
the prediction, while the standard deviation - as expected - more
significantly differs (i.e., $\Delta\widetilde{P}_{RX}^{\mathrm{dev}}\left(A\right)=4.0$
{[}dB{]} vs. $\Delta P_{RX}^{\mathrm{dev}}\left(A\right)=2.36$ {[}dB{]})
due to the non-idealities and inaccuracies in the measurement process.

\noindent As for the \emph{RoI} extension, the experimental tests
indicate that it has been reduced down to $\widetilde{\Lambda}_{SEME}\left(\Omega_{A}\right)=5.50$
{[}$\mathrm{m}^{2}${]} with a (measured) coverage improvement of
$\widetilde{\rho}_{SEME}\left(A\right)=72.7$ \%, which is even better
than the predicted one {[}i.e. $\rho_{SEME}\left(A\right)=70$ \%{]}.
This implies a probability of being below the threshold smaller than
$10$ \% (i.e., $\widetilde{\Theta}_{th}^{SEME}\left(A\right)=8.5\%$
- Fig. 19) starting from $\widetilde{\Theta}_{th}^{Ref}\left(A\right)=30$
\%, while the prediction was more conservative since $\Theta_{th}^{SEME}=18$
\% {[}Fig. 10(\emph{a}){]} being $\Theta_{th}^{Ref}=45$ \%.

\noindent Besides the experimental assessment at the {}``physical''
layer with the measurements of the received power, the effectiveness
of the \emph{SEME} implementation has been also validated from the
\emph{UE} viewpoint. Towards this end, the OOKLA Speedtest$^{\circledR}$
benchmark \cite{OOKLA} has been run on a laptop connected to the
\emph{AP} $\Psi_{A}$ to evaluate the impact of the \emph{SP-EMS}
$\mathcal{S}_{A}$ on the achievable speed in the download and the
upload links. Figure 20 shows the screenshots of such tests at the
instant in which the maximum throughput has been recorded. The \emph{SEME}
implementation enables a remarkable improvement in both the download
{[}$\mathcal{D}_{Ref}^{\max}\left(A\right)=110.12$ {[}Mbps{]} vs.
$\mathcal{D}_{SEME}^{\max}\left(A\right)=131.16$ {[}Mbps{]} - Fig.
20(\emph{a}) vs. Fig. 20(\emph{c}){]} and the upload {[}$\mathcal{U}_{Ref}^{\max}\left(A\right)=84.49$
{[}Mbps{]} vs. $\mathcal{U}_{SEME}^{\max}\left(A\right)=108.19$ {[}Mbps{]}
- Fig. 20(\emph{b}) vs. Fig. 20(\emph{d}){]} links, respectively. 

\noindent For completeness, Table III gives the measured statistics
(minimum, maximum, average, and standard deviation) of both download/upload
throughput ($\mathcal{D}/\mathcal{U}$) and latency ($\mathcal{L}_{\mathcal{D}}/\mathcal{L}_{\mathcal{U}}$).
As it can be noticed, the average download throughput significantly
increases being $\Xi_{\mathcal{D}}\left(A\right)=33.6$ \%, while
the average latency of the download link is more than halved (i.e.,
$\Xi_{\mathcal{L_{D}}}\left(A\right)=-57.6$ \%) {[}$\Xi_{\zeta}\mathcal{\left(\gamma\right)}\triangleq\frac{\zeta_{SEME}^{\mathrm{avg}}\mathcal{\left(\gamma\right)}-\zeta_{Ref}^{\mathrm{avg}}\left(\gamma\right)}{\zeta_{Ref}^{\mathrm{avg}}\left(\gamma\right)}$,
$\zeta\in\left\{ \mathcal{D},\,\mathcal{L}_{\mathcal{D}},\,\mathcal{T}_{\mathcal{D}},\,\mathcal{U},\,\mathcal{L}_{\mathcal{U}}\right\} ${]}.
For instance, this means that the time required to download a $7$
{[}GB{]} 4K-definition movie in the reference scenario would be $\mathcal{T}_{\mathcal{D}}^{Ref}\left(A\right)=9.74$
{[}min{]}, while it would reduce to $\mathcal{T}_{\mathcal{D}}^{SEME}\left(A\right)=7.29$
{[}min{]} in the \emph{SEME} scenario at hand with a time saving of
$\Xi_{\mathcal{T}_{\mathcal{D}}}\left(A\right)=25.2$ \%.

\noindent Similar conclusions hold true in upload, the throughput
and the latency being $\Xi_{\mathcal{U}}\left(A\right)=19.1\%$ and
$\Xi_{\mathcal{L_{U}}}\left(A\right)=-52.9\%$ better than in the
reference scenario (Tab. III).

\subsection{Total Cost of Ownership (\emph{TCO}) Analysis \label{sub:TCO}}

In order to envisage a commercial deployment of the proposed \emph{SEME}
implementation, a comprehensive evaluation of the pros/cons also from
an economic point of view is necessary also besides the numerical/experimental
assessment of its \emph{EM} performance. Towards this end, a preliminary
estimation of the \emph{Total Cost of Ownership} (\emph{TCO}) \cite{Christensen 2016}
has been performed.

\noindent The \emph{TCO}, $\mathbb{C}$, is an index commonly adopted
to estimate the long-term budget for the implementation of a technological
solution for making informed decisions about the investments plan.
As a matter of fact, it allows one to have a feeling on the economic
effectiveness and the sustainability of a new hardware purchase/deployment
devoted to improve the \emph{Wi-Fi} network performance. More specifically,
the \emph{TCO} accounts for different terms related to (\emph{i})
the purchasing ({}``cost of acquisition'', $\mathcal{C}_{a}$),
the installation ({}``cost of commissioning'', $\mathcal{C}_{c}$),
the power consumption ({}``cost of operation'', $\mathcal{C}_{o}$),
and the removal at the end of the life-cycle ({}``cost of decommissioning'',
$\mathcal{C}_{d}$) of a device \cite{Christensen 2016}\begin{equation}
\mathbb{C}=\mathcal{C}_{a}+\mathcal{C}_{c}+\mathcal{C}_{o}+\mathcal{C}_{d}.\label{eq:TCO}\end{equation}
Let us consider the hallway \emph{{}``A}'' as a representative test
case of a class of common spaces in public buildings. The \emph{PCB}
manufacturing of the \emph{SP-EMS} $\mathcal{S}_{A}$ had a cost of
$\mathcal{C}_{a}^{SEME}\left(A\right)=100$ {[}\${]} (Tab. IV),%
\footnote{\noindent Due to the very simple \emph{SP-EMS} layout {[}Fig. 9(\emph{a}){]}
and the exploitation of rather inexpensive dielectric boards, manufacturing
costs are expected to be remarkably lower in a large-scale commercial
production, while here the sample was a prototype realized with a
dedicated production.%
} while the panel installation had an almost negligible impact on the
overall expense (i.e., $\mathcal{C}_{c}^{SEME}\left(A\right)=5$ {[}\${]}
- Tab. IV). Unlike \emph{RP-EMS}s deployments \cite{Trichopoulos 2022},
there are no recurrent costs for the electric energy consumption ($\mathcal{C}_{o}^{SEME}\left(A\right)=0$
{[}\$/year{]} - Tab. IV) thanks to the absence of any biasing circuits.
Moreover, the final decommissioning implies no additional costs since
there is no cabling to remove nor change to the \emph{Wi-Fi} network
infrastructure (i.e., $\mathcal{C}_{d}^{SEME}\left(A\right)=0$ {[}\${]}
- Tab. IV). It turns out that the \emph{TCO} for such a \emph{SEME}
implementation is equal to $\mathbb{C}{}_{SEME}\left(A\right)=105$
{[}\${]} independently on the duration of its life-cycle.

\noindent Differently, a \emph{STD} approach based on the deployment
of an additional (second) \emph{AP}, although more effective in terms
of \emph{EM} radiation (Fig. 10), would imply higher costs. Approximately,
the purchasing, the installation, and the final removal of another
\emph{AP} would need $\mathcal{C}_{a}^{STD}\left(A\right)+\mathcal{C}_{c}^{STD}\left(A\right)+\mathcal{C}_{d}^{STD}\left(A\right)=1100$
{[}\${]}, while the annual cost for the power consumption would amount
to $\mathcal{C}_{o}^{STD}\left(A\right)=50$ {[}\$/year{]} (Tab. IV).

\noindent Consequently, the foreseen overall economic advantage of
the \emph{SEME} solution at hand over the \emph{STD} one in a time
window of $\Delta t=5$ {[}years{]} (i.e., a medium-long term) would
be equal to $\left.\Delta\mathbb{C}\left(\left.A\right|\Delta t\right)\right\rfloor _{\Delta t=5\,[\mathrm{years}]}=1245$
{[}\${]} {[}$\Delta\mathbb{C}\left(\left.A\right|\Delta t\right)\triangleq\mathbb{C}_{STD}\left(\left.A\right|\Delta t\right)-\mathbb{C}_{SEME}\left(A\right)${]}
that is of about $\left.\Xi_{\mathbb{C}}\left(\left.A\right|\Delta t\right)\right\rfloor _{\Delta t=5\,[\mathrm{years}]}\approx92$
\% {[}$\Xi_{\mathbb{C}}\left(\left.A\right|\Delta t\right)\triangleq\frac{\mathbb{C}_{STD}\left(\left.A\right|\Delta t\right)-\mathbb{C}_{SEME}\left(A\right)}{\mathbb{C}_{STD}\left(\left.A\right|\Delta t\right)}${]}.
Even more interestingly, it turns out that the average cost saving
would be of about $\frac{\left.\Delta\mathbb{C}\left(\left.A\right|\Delta t\right)\right\rfloor _{\Delta t=5\,[\mathrm{years}]}}{\Lambda\left(A\right)}\approx18.4$
{[}\$/$\mathrm{m}^{2}${]}. This latter outcome clearly points out
the great potential of such a \emph{SEME} implementation from a practical
perspective.

\section{Conclusions \label{sec:Conclusions}}

\noindent In this work, the first - to the best of the authors' knowledge
- large-scale indoor experimental assessment of an implementation
of the emerging \emph{SEME} paradigm, which is based on the deployment
of \emph{SP-EMS}s realized with low-cost off-the-shelf \emph{PCB}
substrates and not requiring any biasing circuitry, has been documented.
A measurement campaign, aimed at replicating as much as possible a
real-life \emph{UE} by using commodity devices and adopting non-ideal
probing conditions, has been carried out to confirm the performance
predictions drawn from an extensive numerical/tolerance analysis.
The results from both the numerical and the experimental assessments
have proved the following main outcomes:

\begin{itemize}
\item \emph{SP-EMS}s are suitable devices for enhancing the coverage of
large rich-scattering indoor areas without any modification to the
existing \emph{Wi-Fi} network infrastructure, thus enabling a seamless
integration with the existing \emph{AP}s and, more in general, within
the network architecture as well as the buildings;
\item preliminary \emph{RT}-based numerical studies are reliable assets
for faithfully predicting the performance of \emph{SP-EMS}s-based
\emph{SEME} implementations in order to identify the most suitable
layouts/deployments for a successive prototyping/installation;
\item the synthesized \emph{SP-EMS}s exhibit a good robustness to the positioning
errors that potentially occur in real-world deployments;
\item the proposed \emph{SEME} implementation turns out to be economically
advantageous over a traditional (active) coverage-enhancement strategies.
\end{itemize}
Future works will be aimed at implementing commercially-ready \emph{SEME}s
and at developing computationally-efficient strategies for their artificial
intelligence-driven prototyping \cite{Oliveri 2022} and optimal planning
\cite{Salucci 2023}\cite{Benoni 2022} towards a mass-market and
a huge real-world commercial deployment.

\section*{\noindent Acknowledgements}

\noindent \textcolor{black}{This work benefited from the networking
activities carried out within the Project \char`\"{}MITIGO - Mitigazione
dei rischi naturali per la sicurezza e la mobilita' nelle aree montane
del Mezzogiorno\char`\"{} (Grant no. ARS01\_00964) funded by the Italian
Ministry of Education, University, and Research within the PON R\&I
2014-2020 Program (CUP: B64I20000450005) and the Project \char`\"{}SPEED\char`\"{}
(Grant No. 6721001) funded by National Science Foundation of China
under the Chang-Jiang Visiting Professorship Program. A. Massa wishes
to thank E. Vico for her never-ending inspiration, support, guidance,
and help.}

\newpage
\section*{FIGURE CAPTIONS}

\begin{itemize}
\item \textbf{Figure 1.} Picture of the large-scale test-bed situated in
\emph{Mesiano} building of the Department of Civil, Environmental,
and Mechanical Engineering (\emph{DICAM}) of the University of Trento,
Trento (Italy).
\item \textbf{Figure 2.} \emph{Reference Scenario} - Color-map of the (\emph{a})
\emph{RT}-simulated and (\emph{b}) measured received power distributions
at a constant height of $h=1.2$ {[}m{]} (Fig. 3) within the common
places (i.e., hallways {}``\emph{A}'', {}``\emph{B}'', {}``\emph{C}'',
{}``\emph{D}'', and {}``\emph{E}'') of the second floor of the
\emph{Mesiano} building (Fig. 1).
\item \textbf{Figure 3.} Sketch of the measurement geometry.
\item \textbf{Figure 4.} \emph{Reference Scenario} (Hallway {}``\emph{A}'')
- Plot of (\emph{a}) the simulated received power distribution and
(\emph{b}) the corresponding thresholded ($P_{th}=-65$ {[}dBm{]})
binary-map.
\item \textbf{Figure 5.} \emph{Reference Scenario} ($P_{th}=-65$ {[}dBm{]})
- Thresholded binary-map of the (\emph{a}) \emph{RT}-simulated and
(\emph{b}) measured received power distributions within the common
places of the second floor of the \emph{Mesiano} building.
\item \textbf{Figure 6.} Picture of (\emph{a}) the hallway {}``\emph{A}''
where the floor has been marked with paper scotch tape in correspondence
with the uniform sampling grid and (\emph{b}) the hand-held smartphone
used in the measurement campaign.
\item \textbf{Figure 7.} \emph{Reference Scenario} (Hallway {}``\emph{A}'')
- Plot of (\emph{a}) the measured received power distribution and
(\emph{b}) the corresponding thresholded ($P_{th}=-65$ {[}dBm{]})
binary-map.
\item \textbf{Figure 8.} \emph{SEME Scenario} (Hallway \emph{}{}``\emph{A}'')
- \textbf{}Geometrical sketch of the \emph{SP-EMS} deployment: (\emph{a})
global \emph{3-D} view and (\emph{b}) zoom in the area where both
the \emph{AP} ($\Psi_{A}$) and the \emph{SP-EMS} ($\mathcal{S}_{A}$)
are located.
\item \textbf{Figure 9.} Sketch of the layout of the synthesized dual-layers
\emph{SP-EMS}s for (\emph{a}) the hallways {}``\emph{A}'' and {}``\emph{B}''
and (\emph{b}) the hallway {}``\emph{C}''.
\item \textbf{Figure 10.} \emph{SEME Scenario} (Hallway \emph{}{}``\emph{A}'')
- \textbf{}Plot of (\emph{a}) the \emph{CDF} of the received power,
$\Theta$, and (\emph{b}) the \emph{RoI} reduction, $\rho\left(\Omega_{A}\right)$,
versus the \emph{SP-EMS} side, $L_{A}$.
\item \textbf{Figure 11.} \emph{SEME Scenario} (Hallway \emph{}{}``\emph{A}'')
- \textbf{}Full-wave simulated directivity pattern of the \emph{SP-EMS}
$\mathcal{S}_{A}$ in Fig. 9(\emph{a}) of side-length $L_{A}=0.55$
{[}m{]}.
\item \textbf{Figure 12.} \emph{SEME Scenario} (Hallway {}``\emph{A}'',
$L_{A}=0.55$ {[}m{]}) - Plot of (\emph{a}) the simulated received
power distribution, (\emph{b}) the corresponding thresholded ($P_{th}=-65$
{[}dBm{]}) binary-map, and (\emph{c}) the received power improvement
with respect to the {}``reference'' scenario, $\Delta P_{RX}$. 
\item \textbf{Figure 13.} \emph{SEME Scenario} (Hallway {}``\emph{B}'',
$L_{B}=0.55$ {[}m{]}) - Plot of (\emph{a}) the difference map, $\Delta P_{RX}$,
and (\emph{b}) the \emph{CDF} of the received power, $\Theta$. 
\item \textbf{Figure 14.} \emph{SEME Scenario} (Hallway {}``\emph{C}'',
$L_{C}=0.32$ {[}m{]}) - Plot of (\emph{a}) the difference map, $\Delta P_{RX}$,
and (\emph{b}) the \emph{CDF} of the received power, $\Theta$.
\item \textbf{Figure 15.} \emph{SEME Scenario} (\emph{Hallway} {}``\emph{C}'')
- \textbf{}Geometrical sketch of the \emph{SP-EMS} deployment: (\emph{a})
global \emph{3-D} view and (\emph{b}) zoom in the area where both
the \emph{AP} ($\Psi_{C}$) and the \emph{SP-EMS} ($\mathcal{S}_{C}$)
are located.
\item \textbf{Figure 16.} \emph{SEME Scenario} (Hallway {}``\emph{A}'',
$L_{A}=0.55$ {[}m{]}) - \textbf{}Pictures of (\emph{a}) the sketch
of the \emph{EMS} positioning errors, $\Delta y$ and $\Delta z$,
(\emph{b}) the map of the \emph{RoI} reduction, $\rho\left(\Omega_{A}\right)$,
versus $\left(\Delta y,\,\Delta z\right)\in\left[-0.1,\,0.1\right]$
{[}m{]}, (\emph{c}) the plot of the simulated \emph{CDF} in correspondence
with (\emph{c}) horizontal ($\Delta z=0$ {[}m{]}, $\Delta y\in\left[-0.1,\,0.1\right]$
{[}m{]}) or (\emph{d}) vertical ($\Delta y=0$ {[}m{]}, $\Delta z\in\left[-0.1,\,0.1\right]$
{[}m{]}) displacements of the nominal location of the \emph{EMS} $\mathcal{S}_{A}$.
\item \textbf{Figure 17.} \emph{SEME Scenario} (Hallway {}``\emph{A}'',
$L_{A}=0.55$ {[}m{]}) - \textbf{}Picture of (\emph{a}) the fabricated
prototype of the \emph{SP-EMS} layout in Fig. 9(\emph{a}) and (\emph{b})
the real deployment according to Fig. 8.
\item \textbf{Figure 18.} \emph{SEME Scenario} (Hallway {}``\emph{A}'',
$L_{A}=0.55$ {[}m{]}) - \textbf{}Plot of (\emph{a}) the measured
received power distribution and (\emph{b}) the measured received power
improvement with respect to the {}``reference'' scenario, $\Delta P_{RX}$.
\item \textbf{Figure 19.} \emph{SEME Scenario} (Hallway {}``\emph{A}'',
$L_{A}=0.55$ {[}m{]}) - \textbf{}Plot of both the simulated and the
measured \emph{CDF}s of the received power (i.e., $\Theta$ and $\widetilde{\Theta}$).
\item \textbf{Figure 20.} Screenshot of the OOKLA Speedtest$^{\circledR}$
\cite{OOKLA} for (\emph{a})(\emph{c}) the download and (\emph{b})(\emph{d})
the upload links in (\emph{a})(\emph{b}) the {}``reference'' scenario
and (\emph{c})(\emph{d}) the \emph{SEME} ($L_{A}=0.55$ {[}m{]}) one
at the same throughput-peak time-instant.
\end{itemize}

\section*{TABLE CAPTIONS}

\begin{itemize}
\item \textbf{Table I.} \emph{SEME Scenario} (Hallway {}``\emph{A}'')
- \textbf{}Coverage-performance indexes in correspondence with different
side-lengths of the \emph{SP-EMS} $\mathcal{S}_{A}$ in Fig. 9(\emph{a}).
\item \textbf{Table II.} \emph{SEME Scenario} (Hallways {}``\emph{B}'',
{}``\emph{C}'', {}``\emph{D}'', {}``\emph{E}'') - Coverage-performance
indexes.
\item \textbf{Table III.} OOKLA Speedtest$^{\circledR}$ \cite{OOKLA} statistics.
\item \textbf{Table IV.} \emph{TCO}-analysis indexes.
\end{itemize}
\newpage
\begin{center}~\vfill\end{center}

\begin{center}\includegraphics[%
  width=0.90\columnwidth]{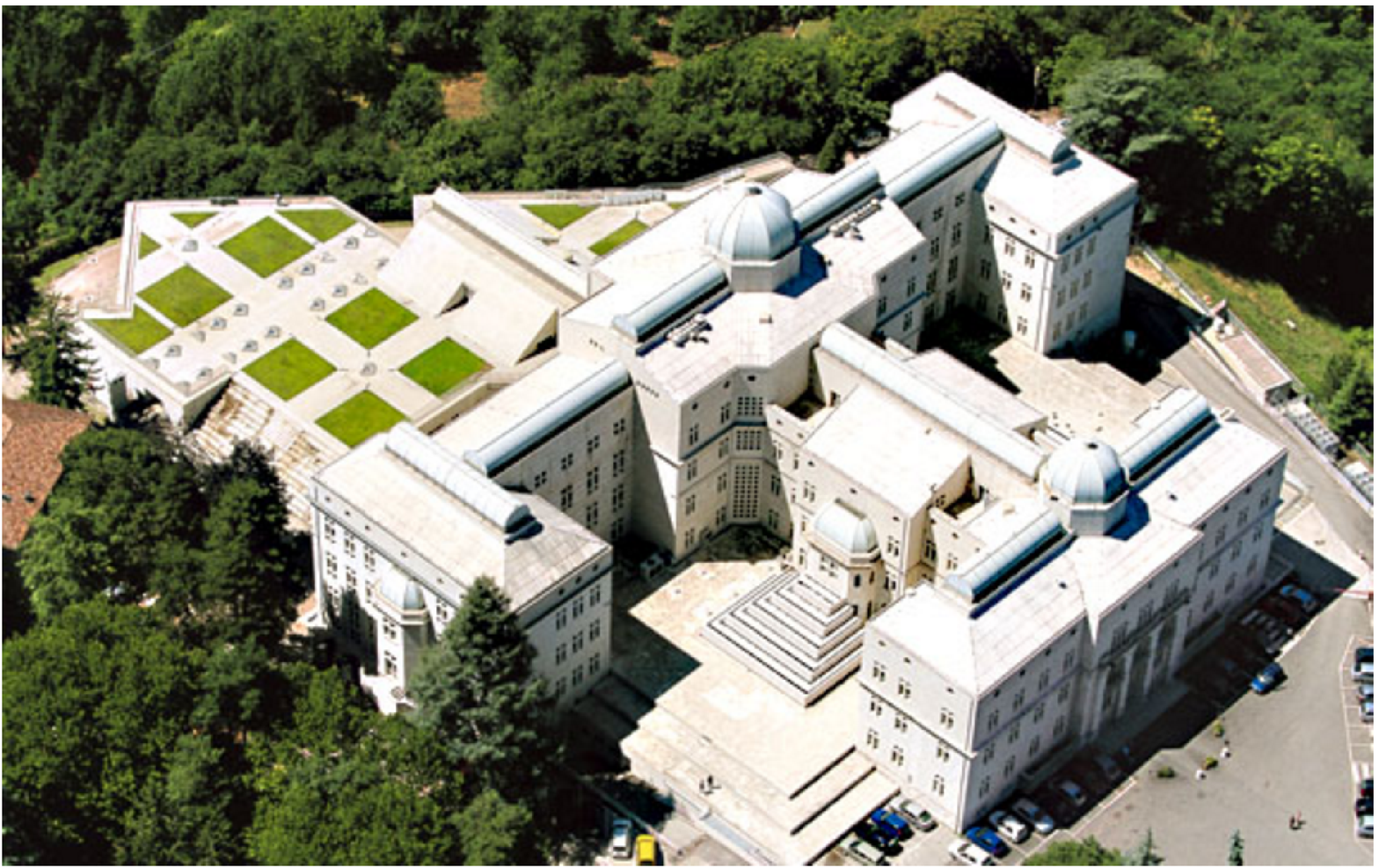}\end{center}

\begin{center}~\vfill\end{center}

\begin{center}\textbf{Fig. 1 - A. Benoni et} \textbf{\emph{al.}}\textbf{,}
\textbf{\emph{{}``}}Towards Real-World Indoor Smart ...''\end{center}

\newpage
\begin{center}~\vfill\end{center}

\begin{center}\begin{tabular}{c}
\includegraphics[%
  width=0.90\columnwidth]{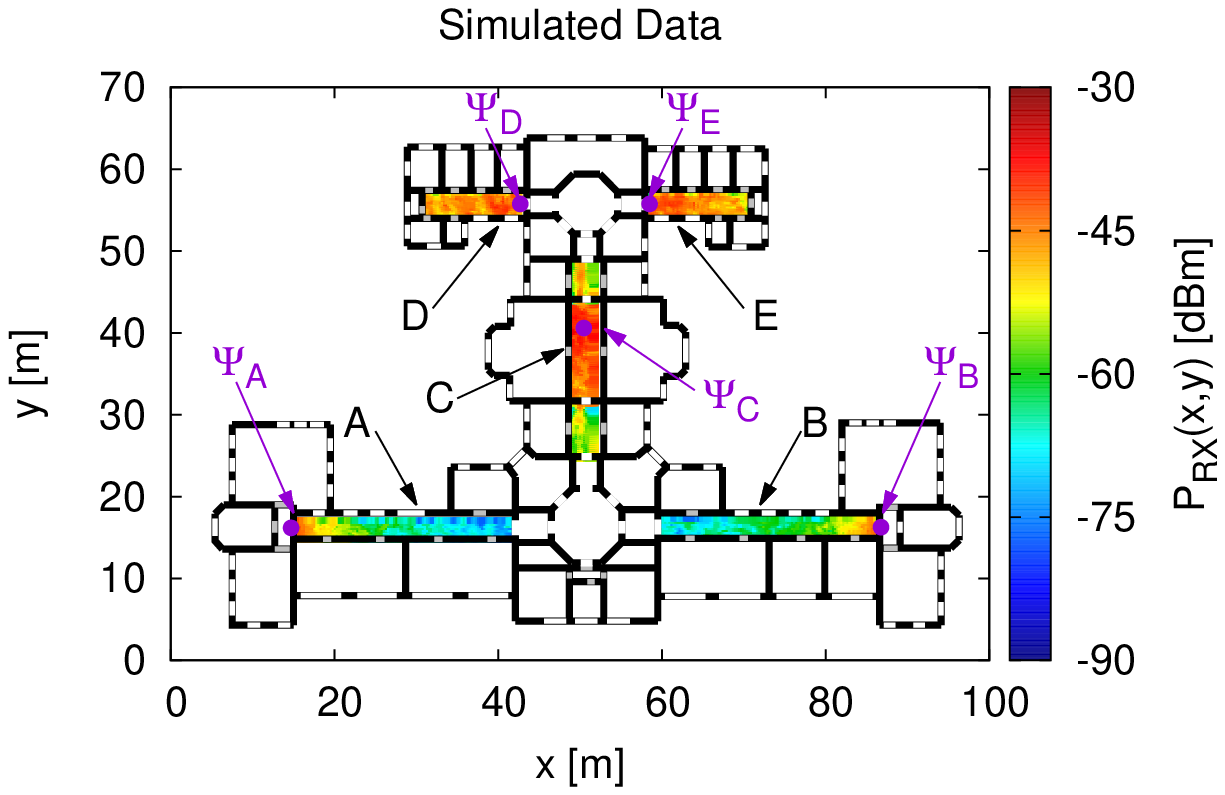}\tabularnewline
(\emph{a})\tabularnewline
\tabularnewline
\includegraphics[%
  width=0.90\columnwidth]{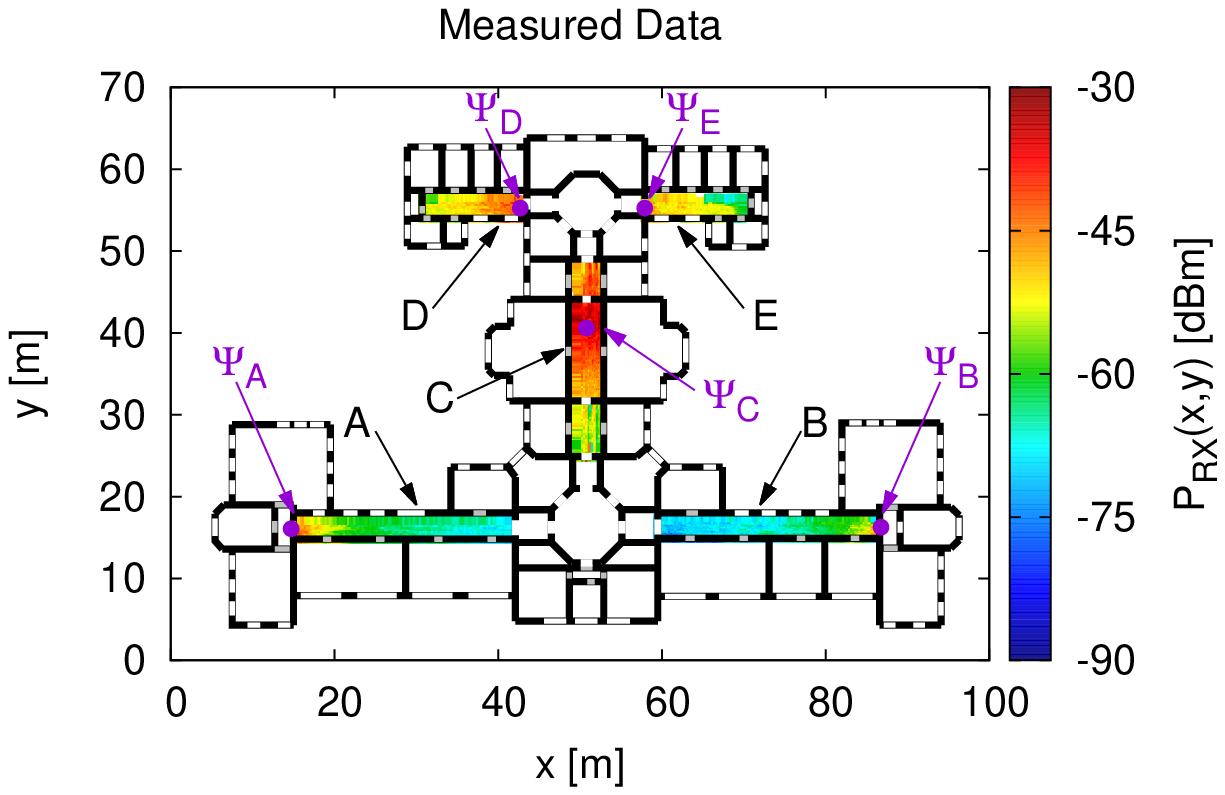}\tabularnewline
(\emph{b})\tabularnewline
\end{tabular}\end{center}

\begin{center}~\vfill\end{center}

\begin{center}\textbf{Fig. 2 - A. Benoni et} \textbf{\emph{al.}}\textbf{,}
\textbf{\emph{{}``}}Towards Real-World Indoor Smart ...''\end{center}

\newpage
\begin{center}~\vfill\end{center}

\begin{center}\begin{tabular}{c}
\includegraphics[%
  width=0.70\columnwidth]{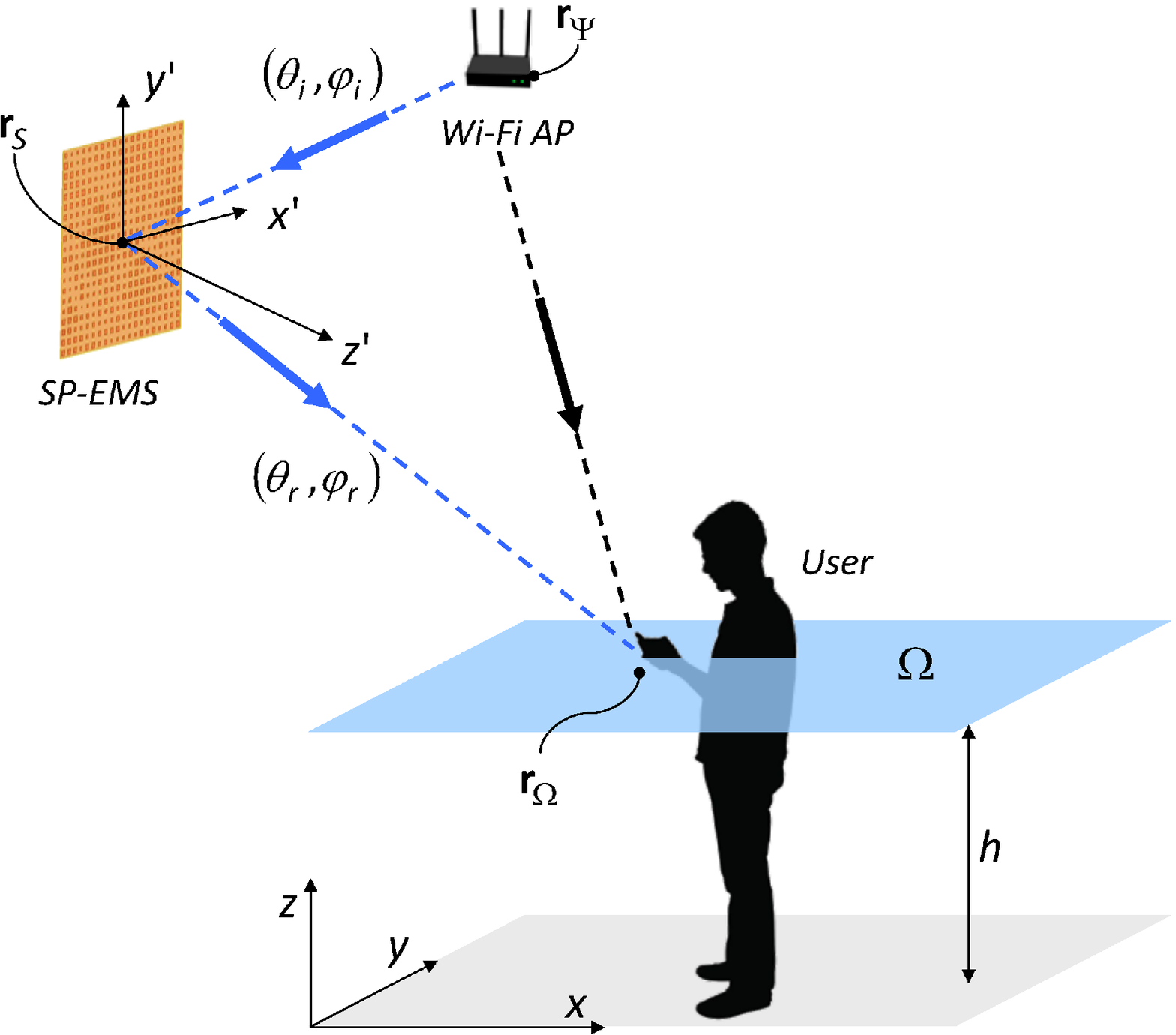}\tabularnewline
\end{tabular}\end{center}

\begin{center}~\vfill\end{center}

\begin{center}\textbf{Fig. 3 - A. Benoni et} \textbf{\emph{al.}}\textbf{,}
\textbf{\emph{{}``}}Towards Real-World Indoor Smart ...''\end{center}

\newpage
\begin{center}~\vfill\end{center}

\begin{center}\begin{tabular}{c}
\includegraphics[%
  width=1.0\columnwidth]{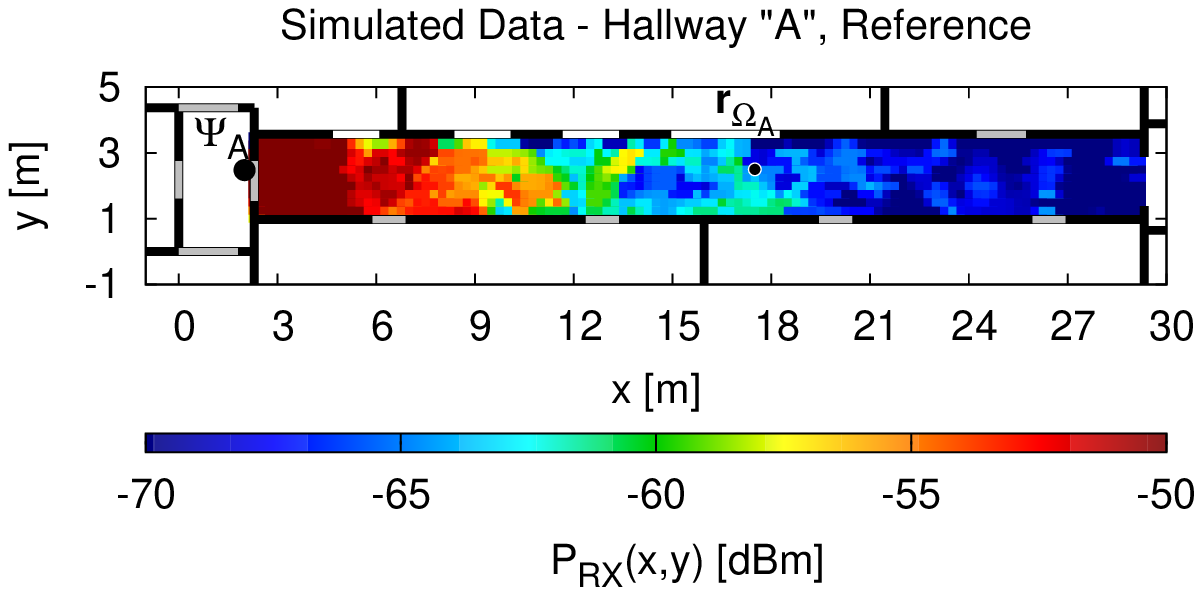}\tabularnewline
(\emph{a})\tabularnewline
\tabularnewline
\includegraphics[%
  width=1.0\columnwidth]{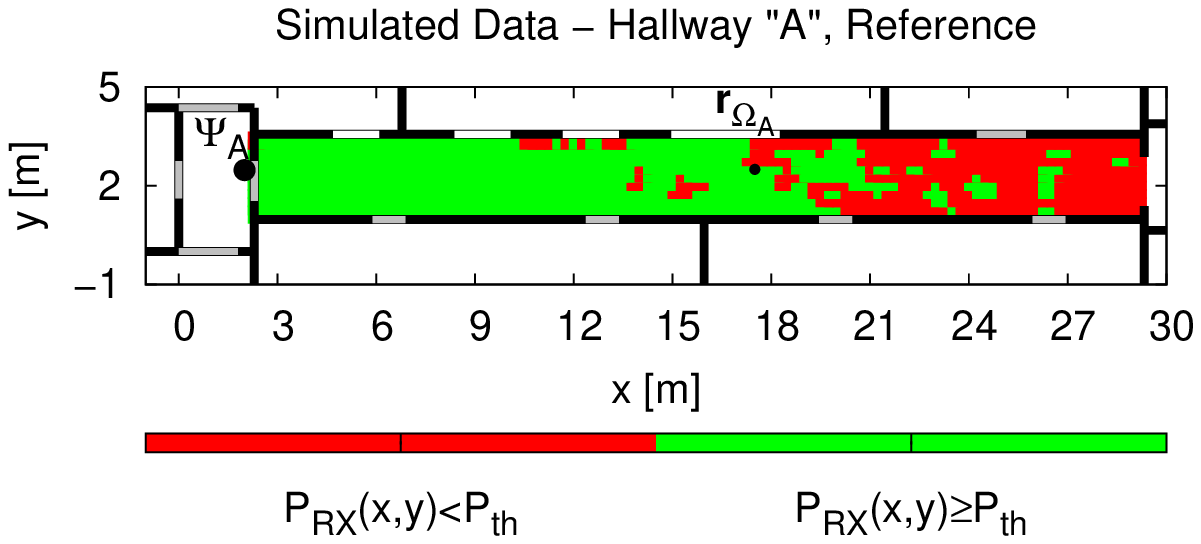}\tabularnewline
(\emph{b})\tabularnewline
\end{tabular}\end{center}

\begin{center}~\vfill\end{center}

\begin{center}\textbf{Fig. 4 - A. Benoni et} \textbf{\emph{al.}}\textbf{,}
\textbf{\emph{{}``}}Towards Real-World Indoor Smart ...''\end{center}

\newpage
\begin{center}~\vfill\end{center}

\begin{center}\begin{tabular}{c}
\includegraphics[%
  width=0.90\columnwidth]{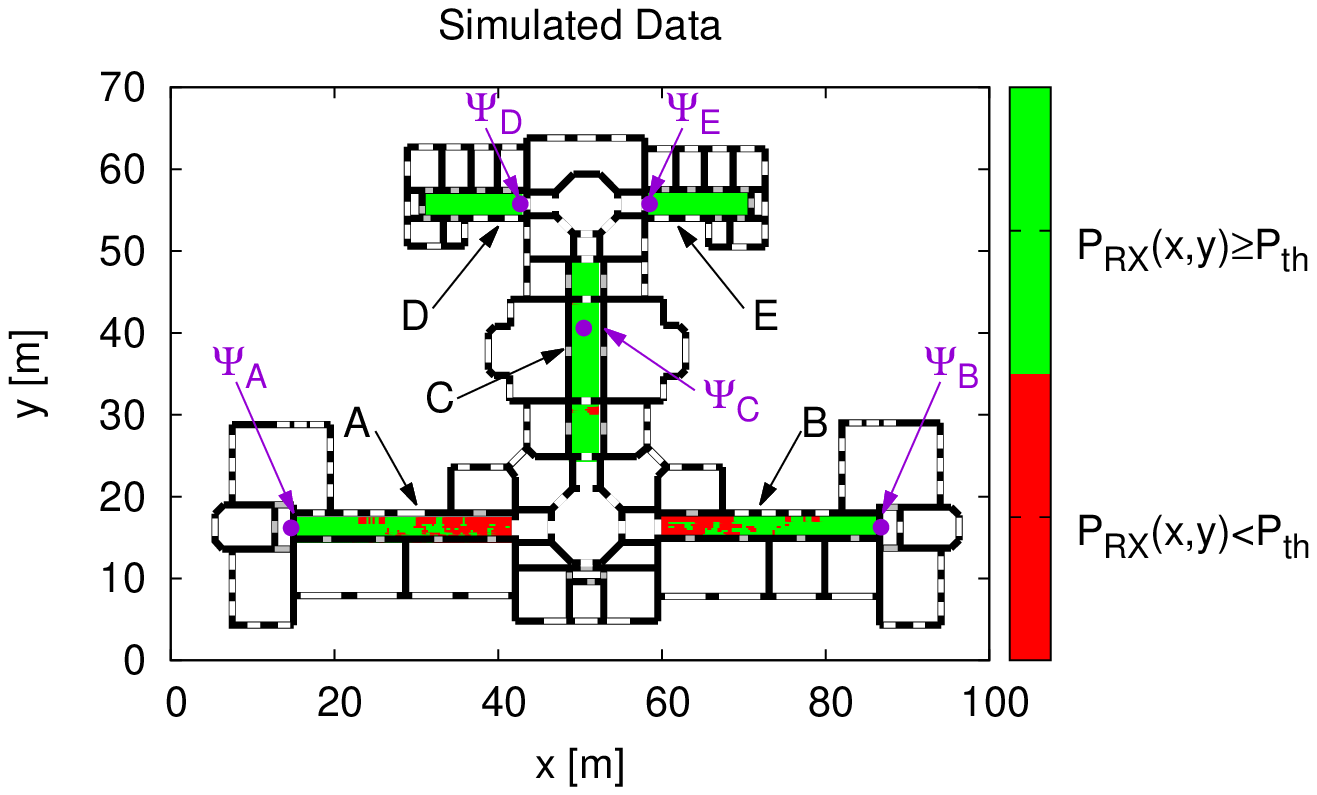}\tabularnewline
(\emph{a})\tabularnewline
\tabularnewline
\includegraphics[%
  width=0.90\columnwidth]{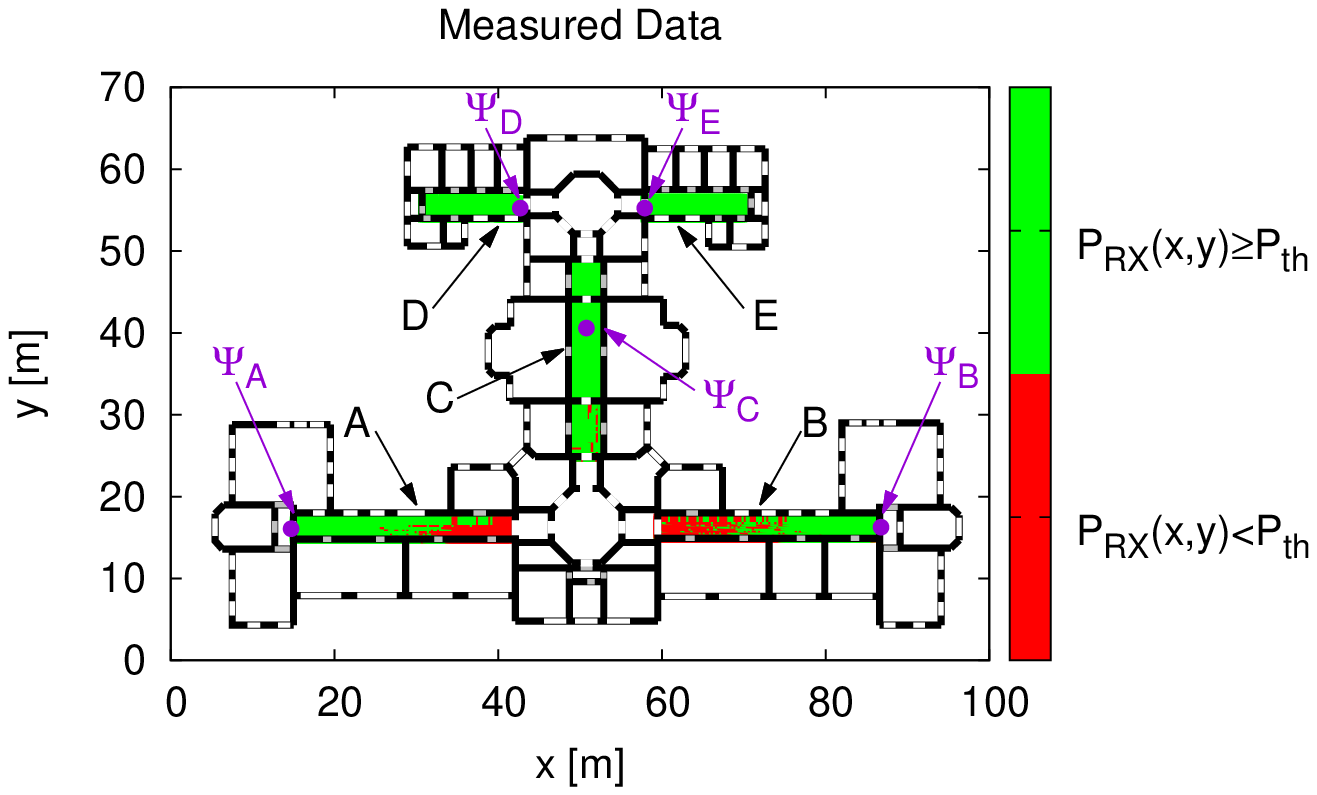}\tabularnewline
(\emph{b})\tabularnewline
\end{tabular}\end{center}

\begin{center}~\vfill\end{center}

\begin{center}\textbf{Fig. 5 - A. Benoni et} \textbf{\emph{al.}}\textbf{,}
\textbf{\emph{{}``}}Towards Real-World Indoor Smart ...''\end{center}

\newpage
\begin{center}~\vfill\end{center}

\begin{center}\begin{tabular}{c}
\includegraphics[%
  width=0.55\columnwidth]{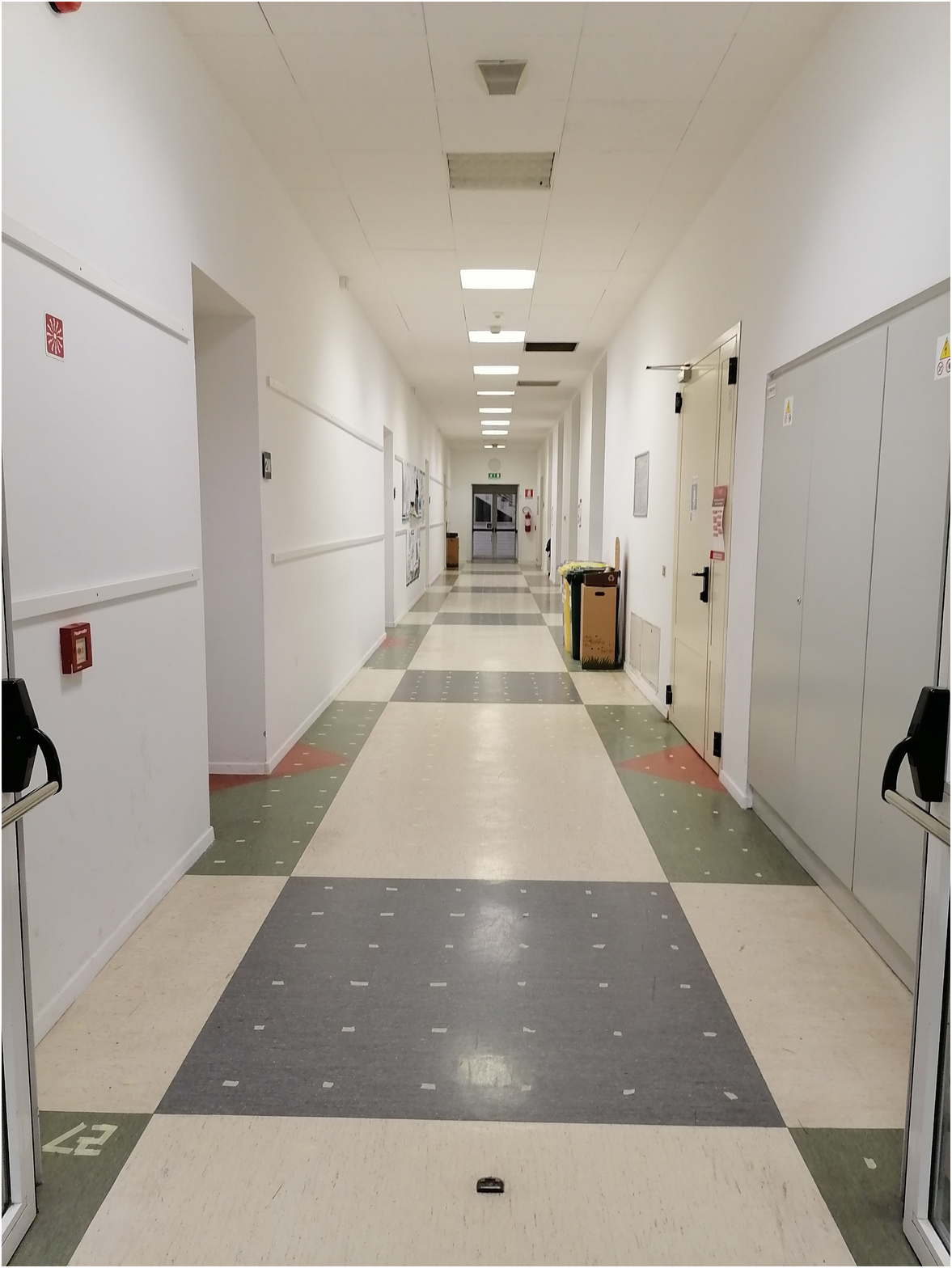}\tabularnewline
(\emph{a})\tabularnewline
\tabularnewline
\includegraphics[%
  width=0.75\columnwidth]{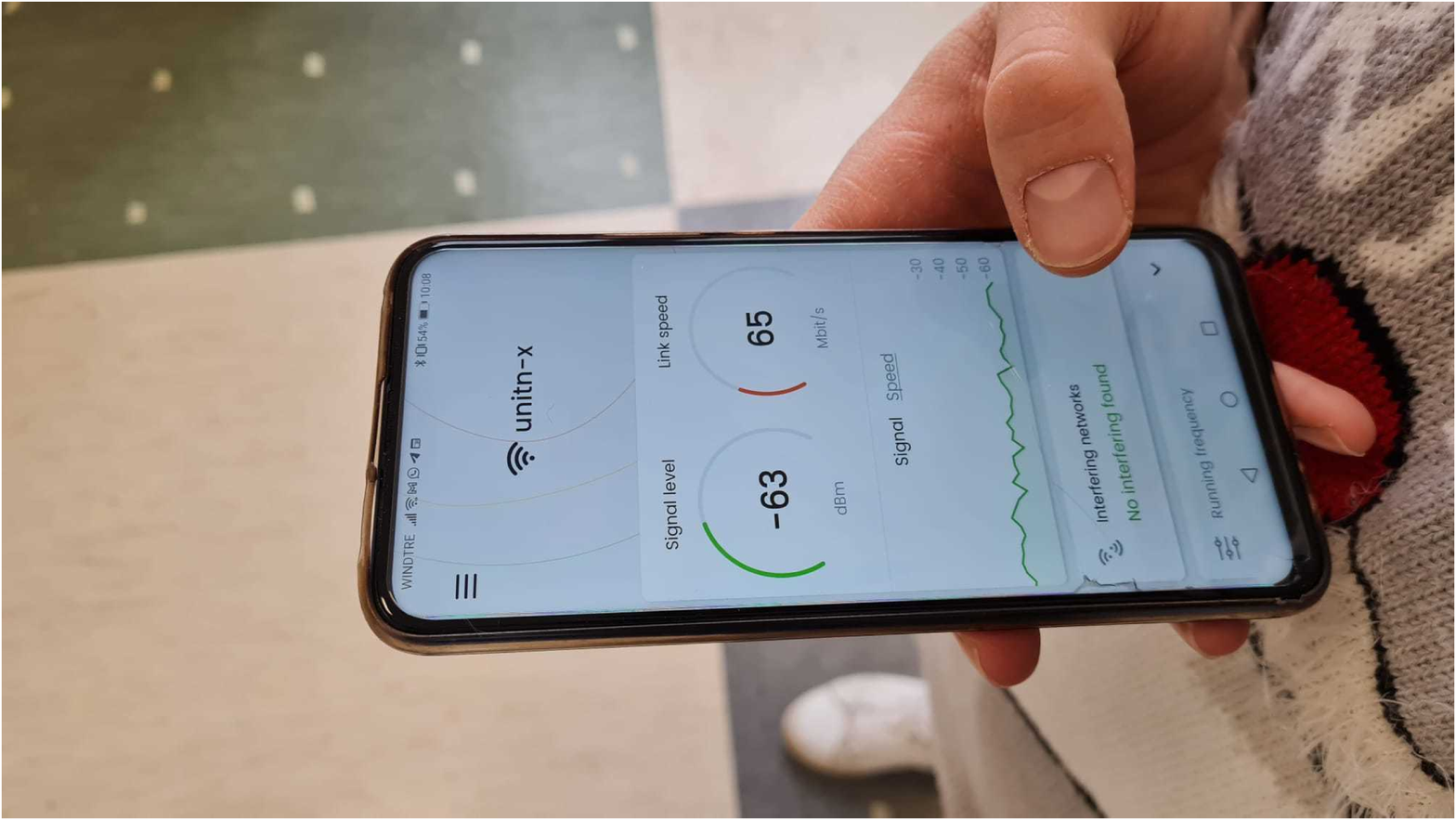}\tabularnewline
(\emph{b})\tabularnewline
\end{tabular}\end{center}

\begin{center}~\vfill\end{center}

\begin{center}\textbf{Fig. 6 - A. Benoni et} \textbf{\emph{al.}}\textbf{,}
\textbf{\emph{{}``}}Towards Real-World Indoor Smart ...''\end{center}

\newpage
\begin{center}~\vfill\end{center}

\begin{center}\begin{tabular}{c}
\includegraphics[%
  width=0.90\columnwidth]{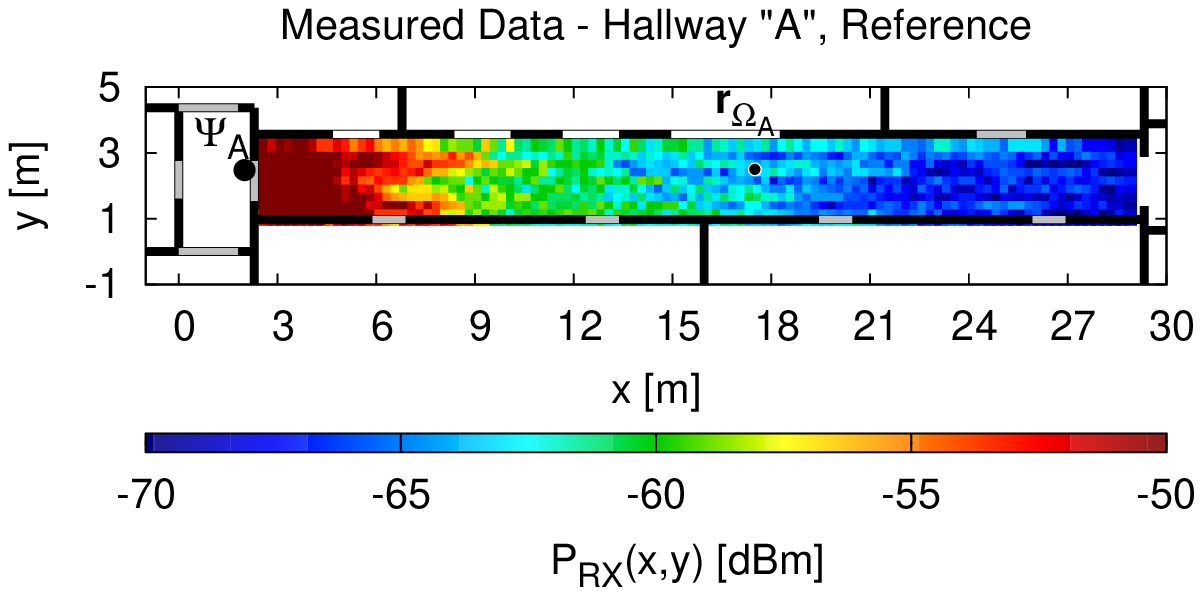}\tabularnewline
(\emph{a})\tabularnewline
\tabularnewline
\includegraphics[%
  width=0.90\columnwidth]{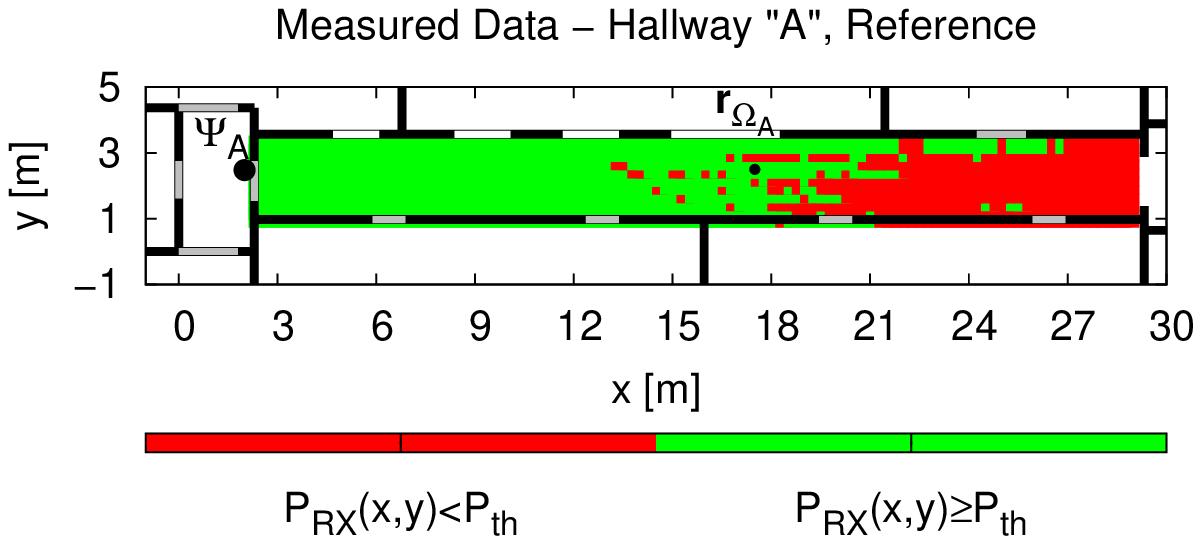}\tabularnewline
(\emph{b})\tabularnewline
\end{tabular}\end{center}

\begin{center}~\vfill\end{center}

\begin{center}\textbf{Fig. 7 - A. Benoni et} \textbf{\emph{al.}}\textbf{,}
\textbf{\emph{{}``}}Towards Real-World Indoor Smart ...''\end{center}

\newpage
\begin{center}~\vfill\end{center}

\begin{center}\begin{tabular}{cc}
\multicolumn{2}{c}{\includegraphics[%
  width=0.70\columnwidth]{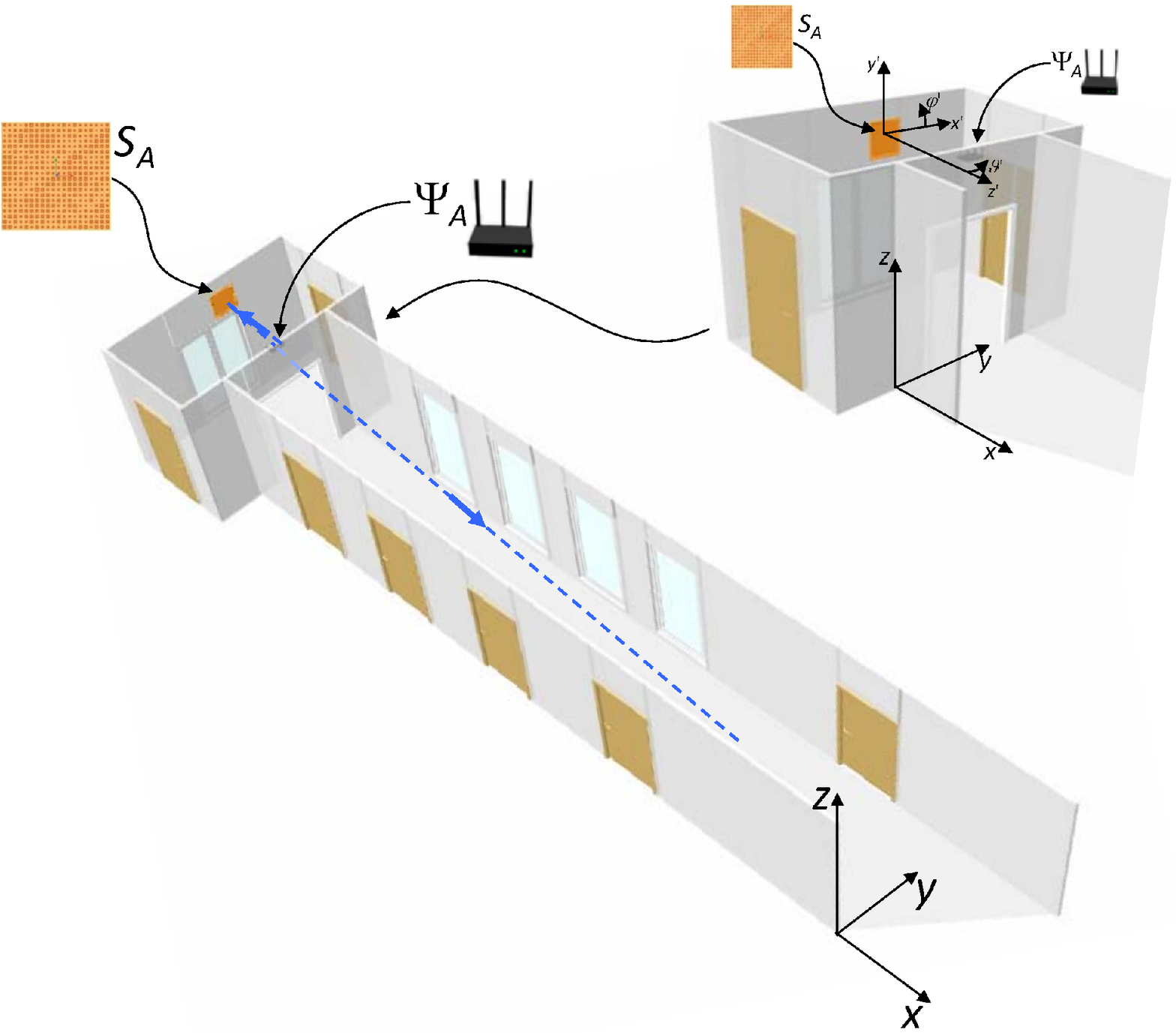}}\tabularnewline
\multicolumn{2}{c}{(\emph{a})}\tabularnewline
\multicolumn{2}{c}{}\tabularnewline
\multicolumn{2}{c}{\includegraphics[%
  width=0.60\columnwidth]{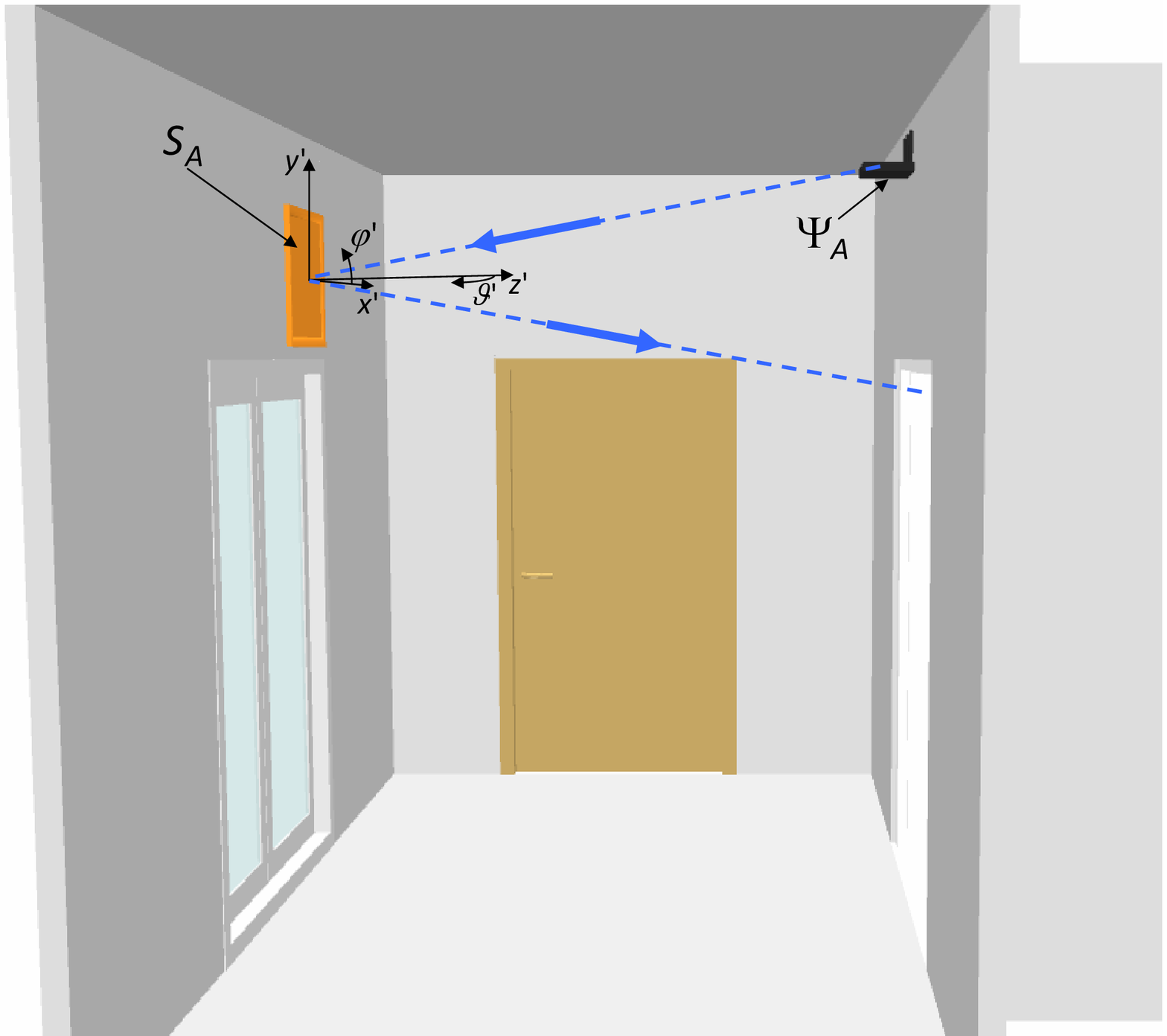}}\tabularnewline
\multicolumn{2}{c}{(\emph{b})}\tabularnewline
\end{tabular}\end{center}

\begin{center}~\vfill\end{center}

\begin{center}\textbf{Fig. 8 - A. Benoni et} \textbf{\emph{al.}}\textbf{,}
\textbf{\emph{{}``}}Towards Real-World Indoor Smart ...''\end{center}

\newpage
\begin{center}~\vfill\end{center}

\begin{center}\begin{tabular}{c}
\includegraphics[%
  width=0.70\columnwidth]{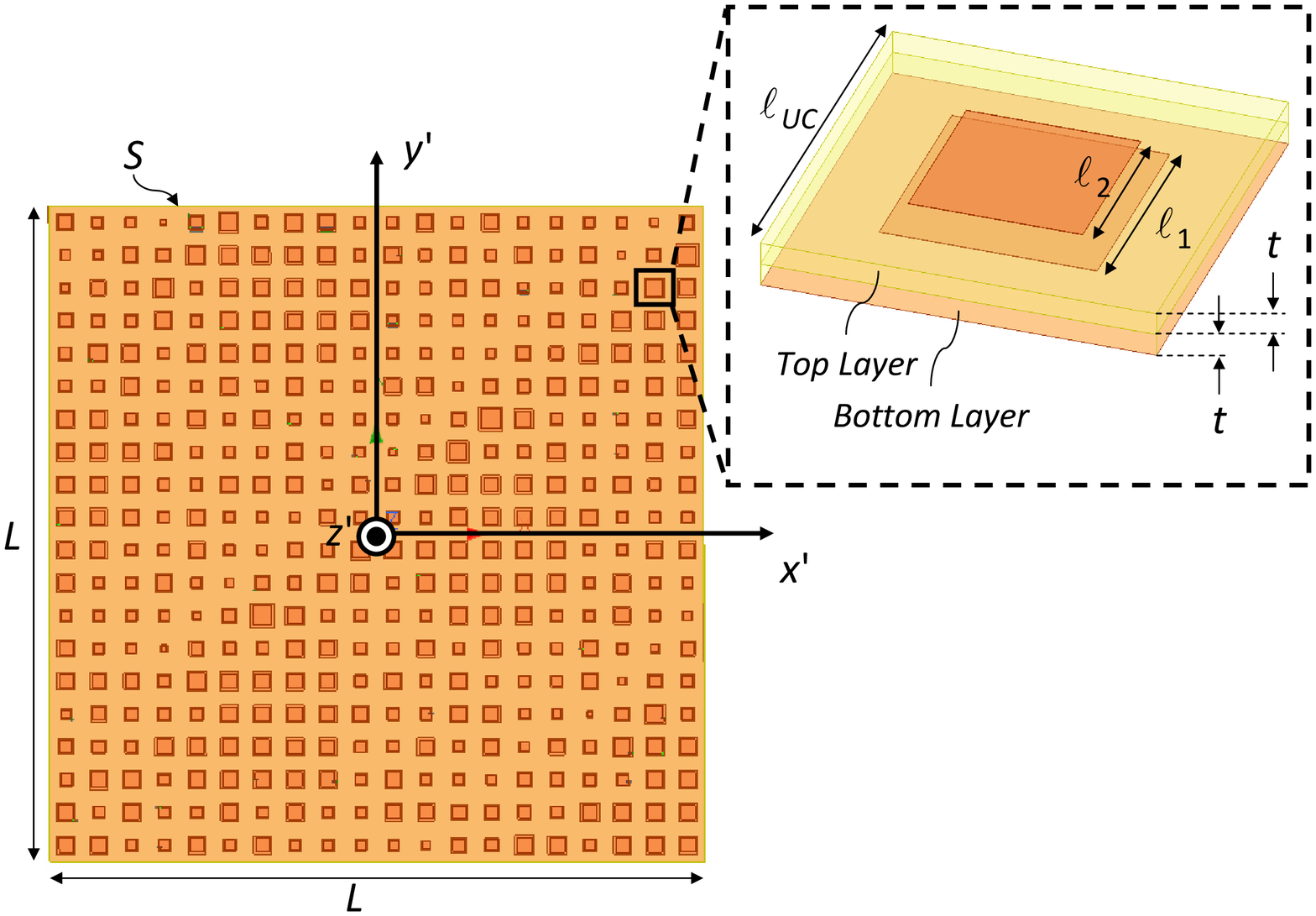}\tabularnewline
(\emph{a})\tabularnewline
\tabularnewline
\includegraphics[%
  width=0.35\columnwidth]{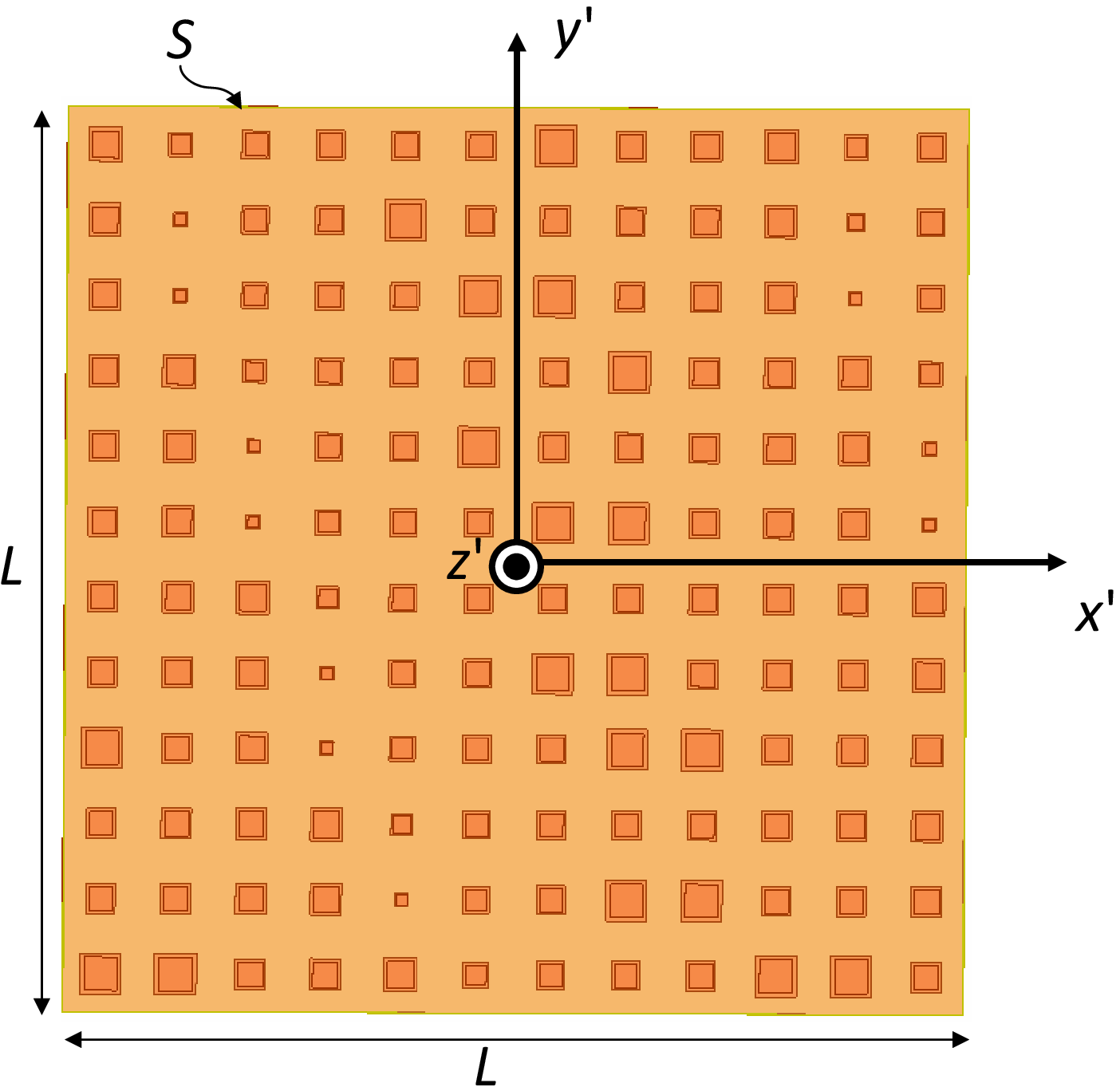}\tabularnewline
(\emph{b})\tabularnewline
\end{tabular}\end{center}

\begin{center}~\vfill\end{center}

\begin{center}\textbf{Fig. 9 - A. Benoni et} \textbf{\emph{al.}}\textbf{,}
\textbf{\emph{{}``}}Towards Real-World Indoor Smart ...''\end{center}

\newpage
\begin{center}~\vfill\end{center}

\begin{center}\begin{tabular}{c}
\includegraphics[%
  width=0.75\columnwidth]{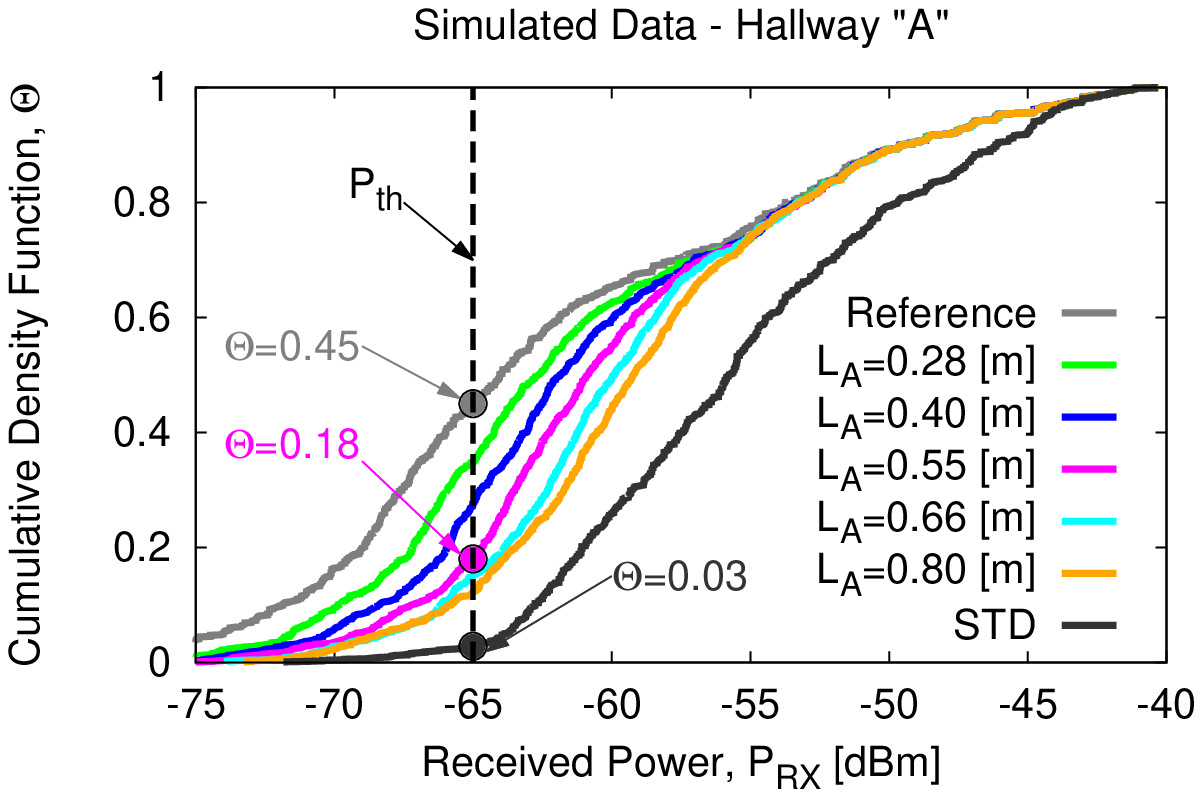}\tabularnewline
(\emph{a})\tabularnewline
\tabularnewline
\includegraphics[%
  width=0.75\columnwidth]{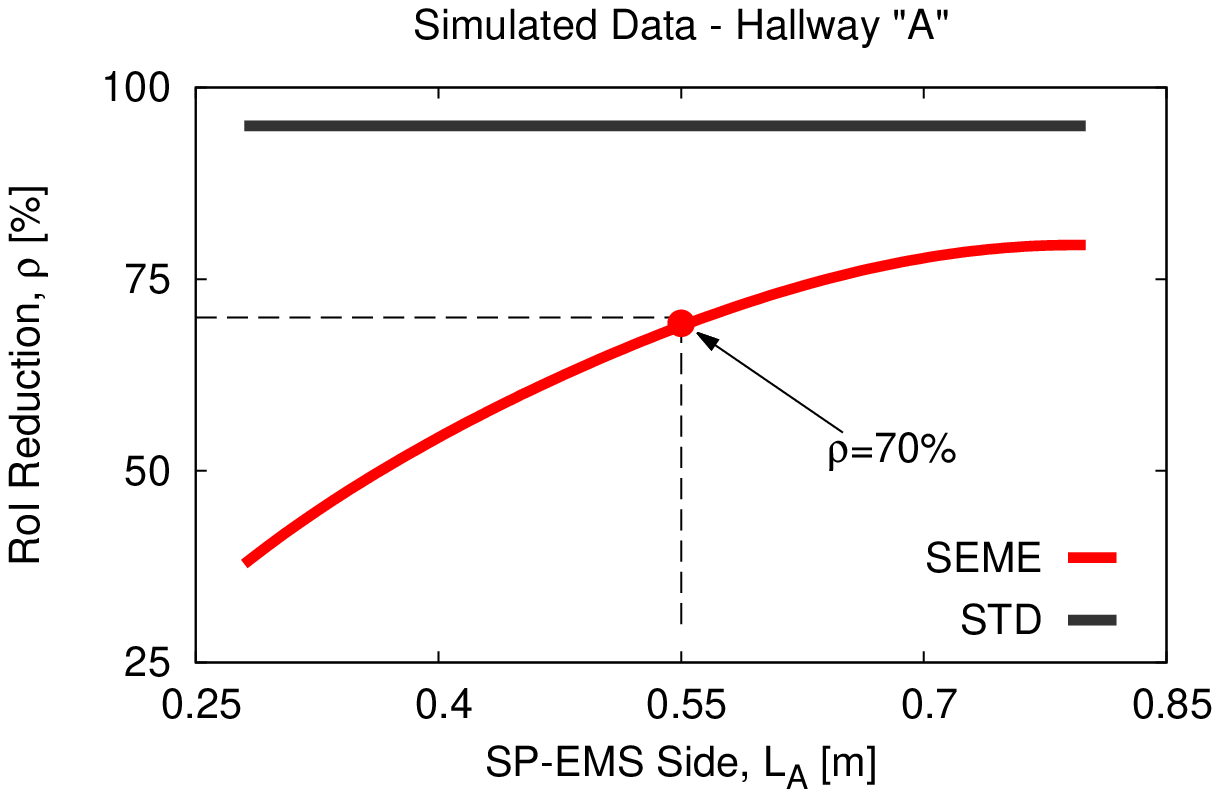}\tabularnewline
(\emph{b})\tabularnewline
\end{tabular}\end{center}

\begin{center}~\vfill\end{center}

\begin{center}\textbf{Fig. 10 - A. Benoni et} \textbf{\emph{al.}}\textbf{,}
\textbf{\emph{{}``}}Towards Real-World Indoor Smart ...''\end{center}

\newpage
\begin{center}~\vfill\end{center}

\begin{center}\includegraphics[%
  width=0.70\columnwidth]{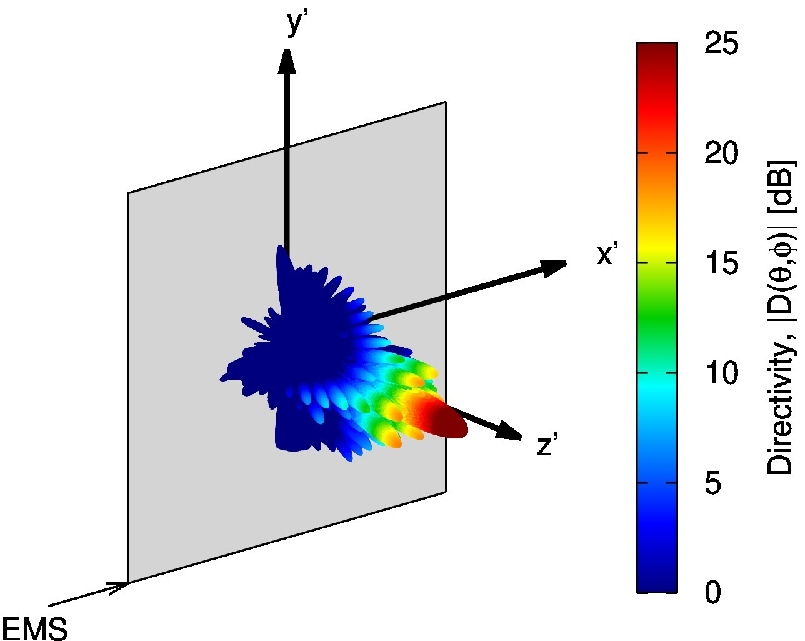}\end{center}

\begin{center}~\vfill\end{center}

\begin{center}\textbf{Fig. 11 - A. Benoni et} \textbf{\emph{al.}}\textbf{,}
\textbf{\emph{{}``}}Towards Real-World Indoor Smart ...''\end{center}

\newpage
\begin{center}~\vfill\end{center}

\begin{center}\begin{tabular}{c}
\includegraphics[%
  width=0.72\columnwidth]{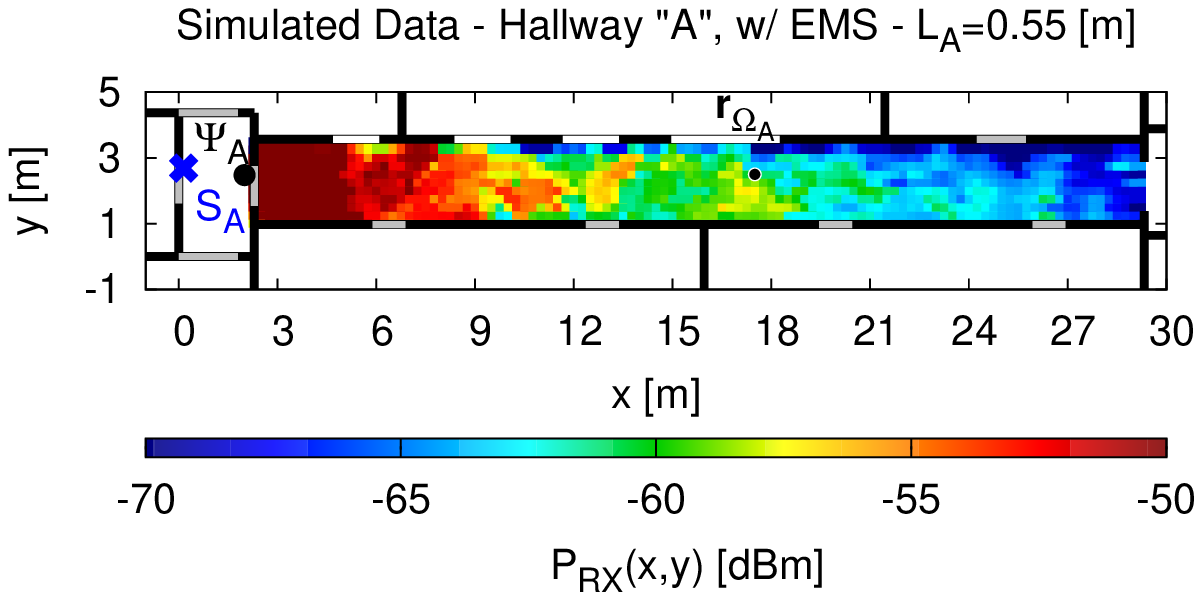}\tabularnewline
(\emph{a})\tabularnewline
\tabularnewline
\includegraphics[%
  width=0.72\columnwidth]{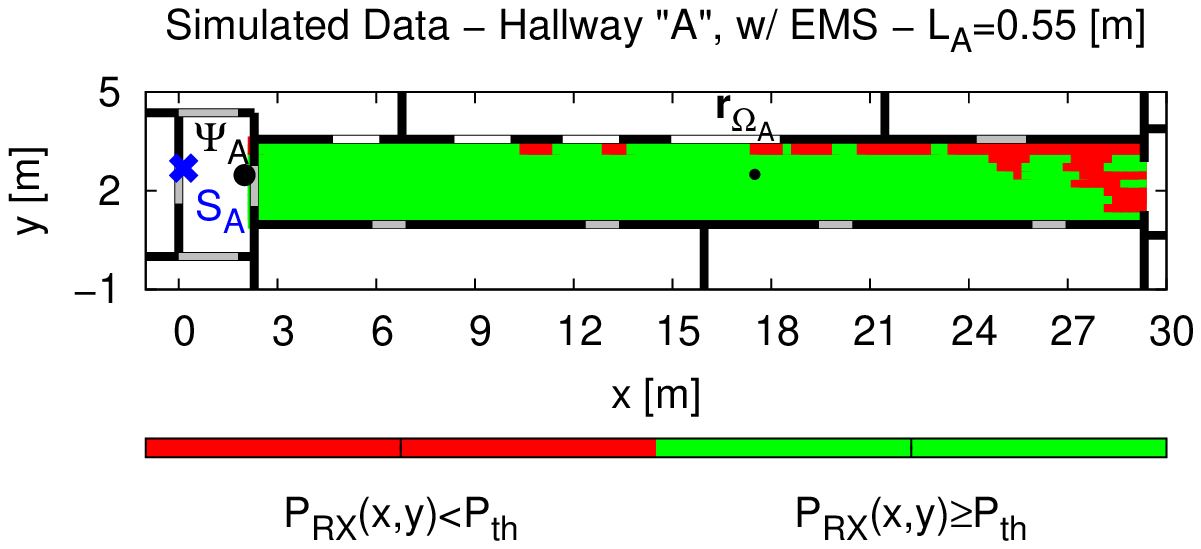}\tabularnewline
(\emph{b})\tabularnewline
\tabularnewline
\includegraphics[%
  width=0.72\columnwidth]{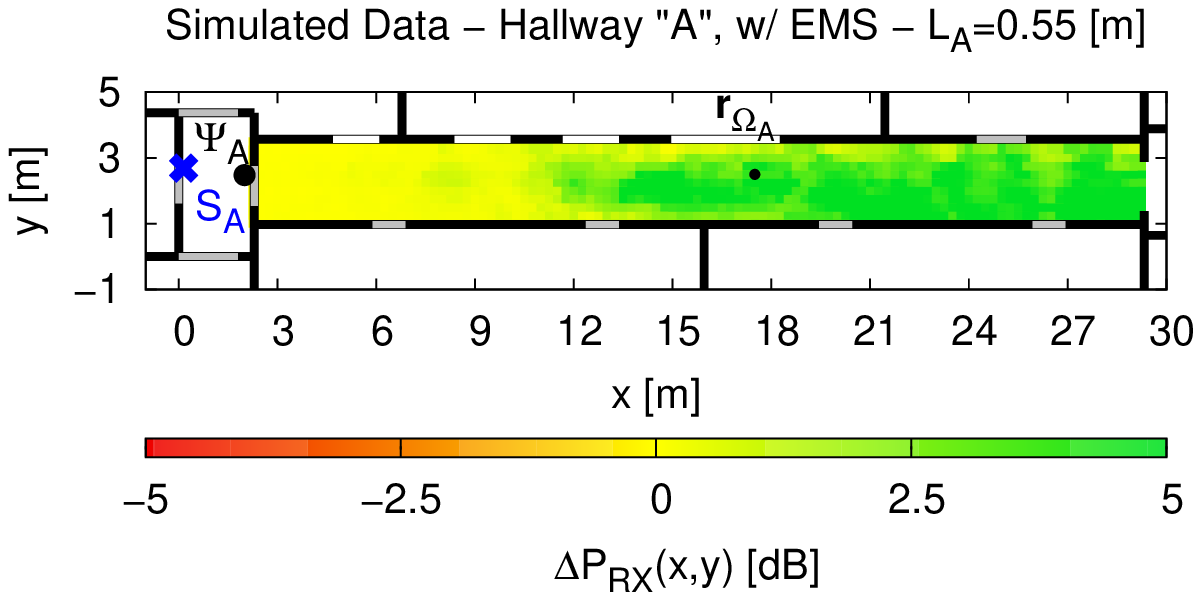}\tabularnewline
(\emph{c})\tabularnewline
\end{tabular}\end{center}

\begin{center}~\vfill\end{center}

\begin{center}\textbf{Fig. 12 - A. Benoni et} \textbf{\emph{al.}}\textbf{,}
\textbf{\emph{{}``}}Towards Real-World Indoor Smart ...''\end{center}

\newpage
\begin{center}~\vfill\end{center}

\begin{center}\begin{tabular}{c}
\includegraphics[%
  width=0.75\columnwidth]{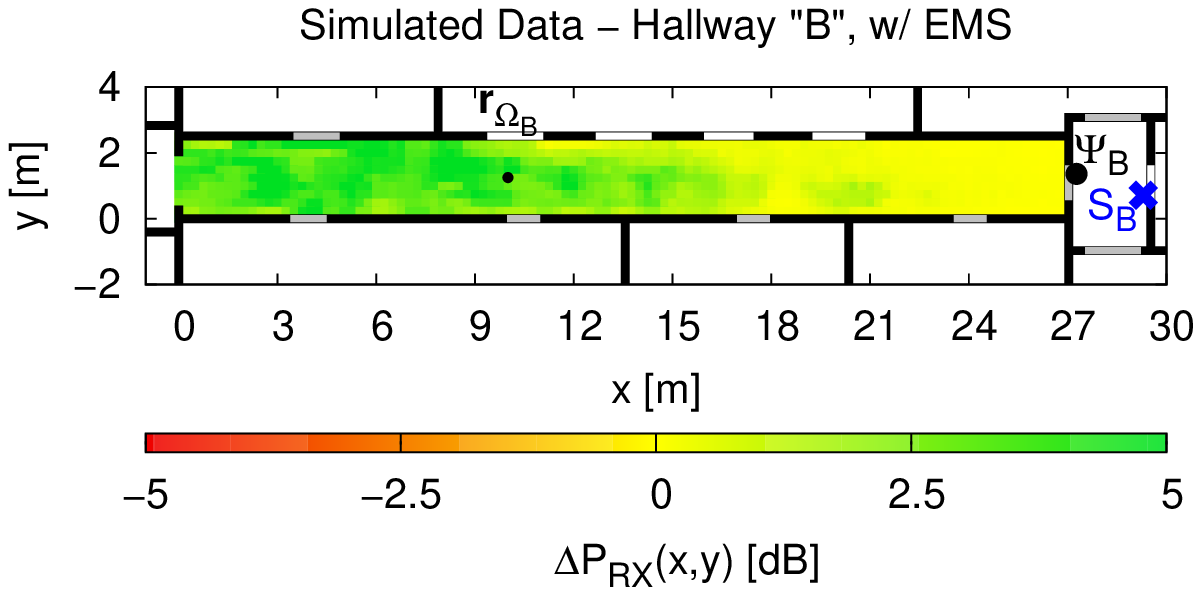}\tabularnewline
(\emph{a})\tabularnewline
\tabularnewline
\includegraphics[%
  width=0.75\columnwidth]{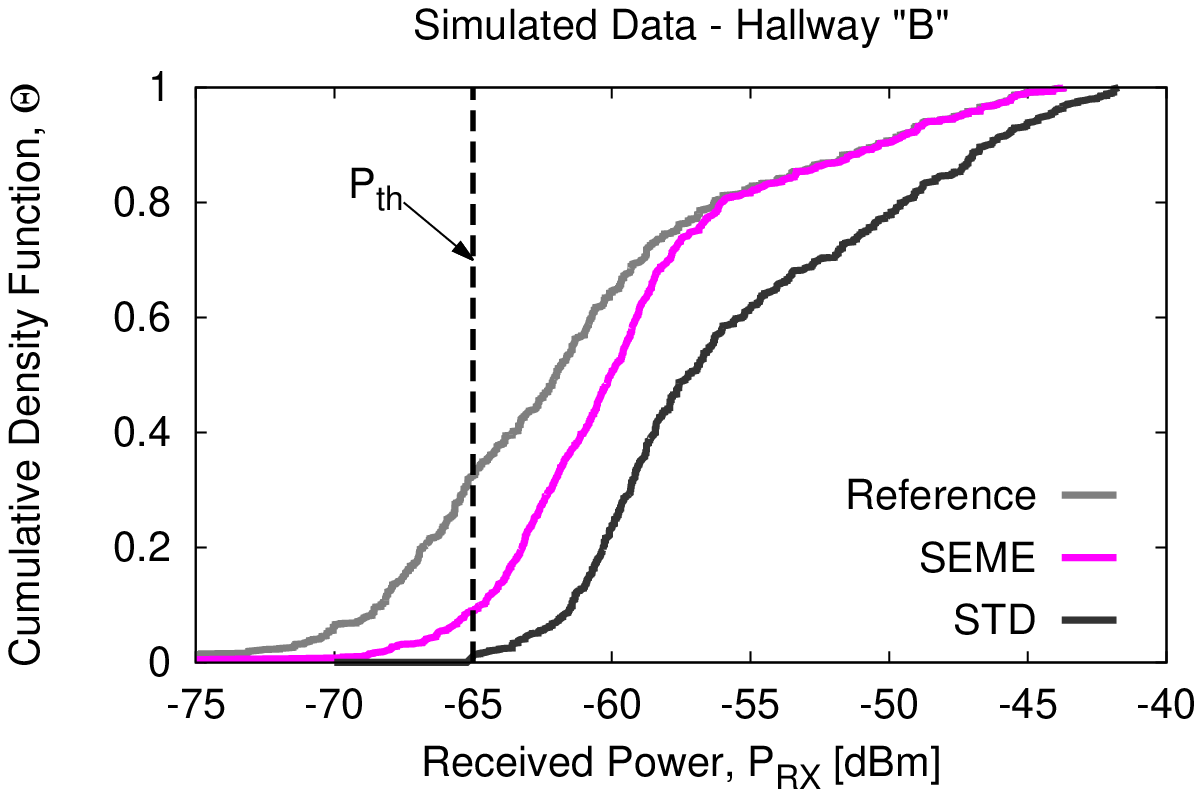}\tabularnewline
(\emph{b})\tabularnewline
\end{tabular}\end{center}

\begin{center}~\vfill\end{center}

\begin{center}\textbf{Fig. 13 - A. Benoni et} \textbf{\emph{al.}}\textbf{,}
\textbf{\emph{{}``}}Towards Real-World Indoor Smart ...''\end{center}

\newpage
\begin{center}~\vfill\end{center}

\begin{center}\begin{tabular}{c}
\includegraphics[%
  width=0.50\columnwidth]{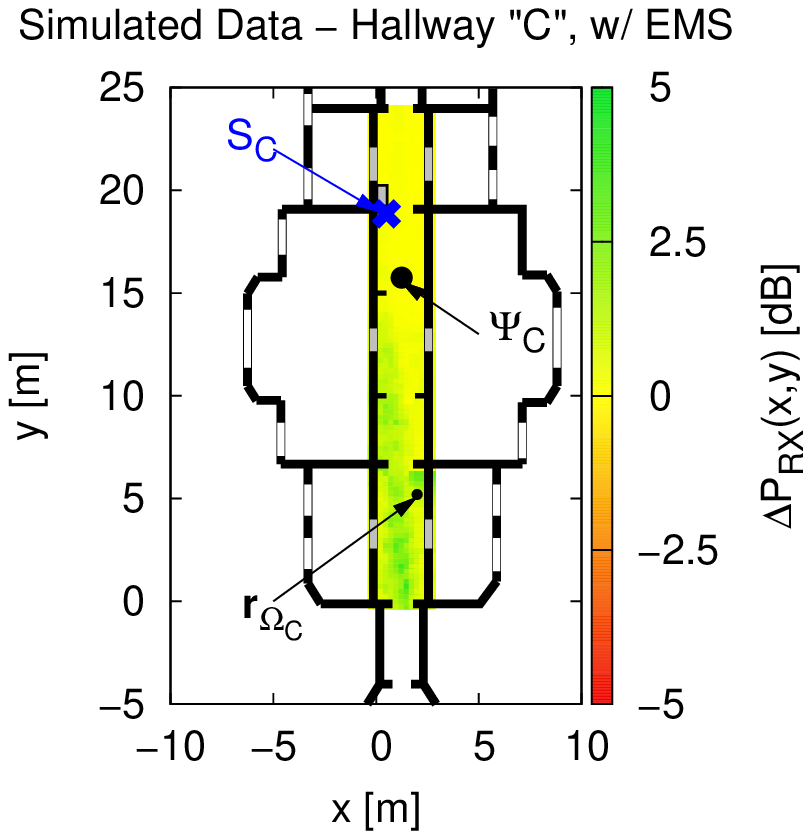}\tabularnewline
(\emph{a})\tabularnewline
\tabularnewline
\includegraphics[%
  width=0.75\columnwidth]{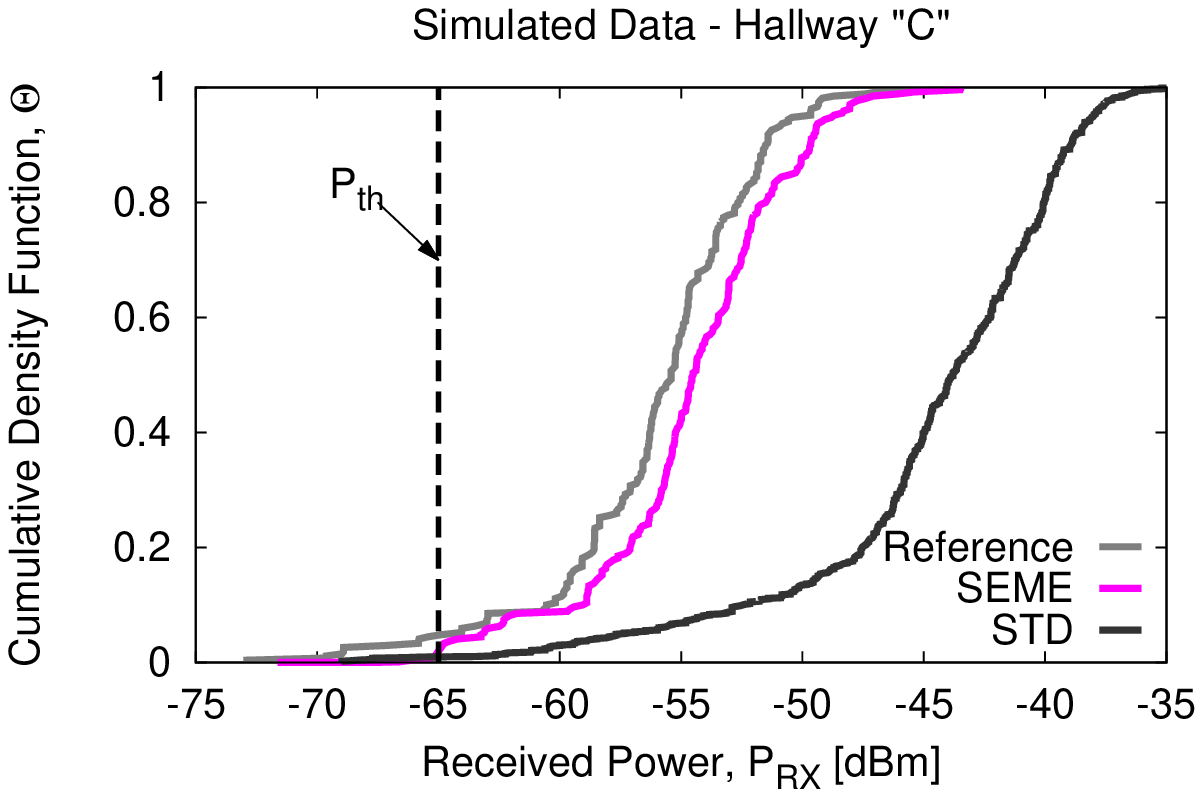}\tabularnewline
(\emph{b})\tabularnewline
\end{tabular}\end{center}

\begin{center}~\vfill\end{center}

\begin{center}\textbf{Fig. 14 - A. Benoni et} \textbf{\emph{al.}}\textbf{,}
\textbf{\emph{{}``}}Towards Real-World Indoor Smart ...''\end{center}

\newpage
\begin{center}~\vfill\end{center}

\begin{center}\begin{tabular}{cc}
\multicolumn{2}{c}{\includegraphics[%
  width=1.0\columnwidth]{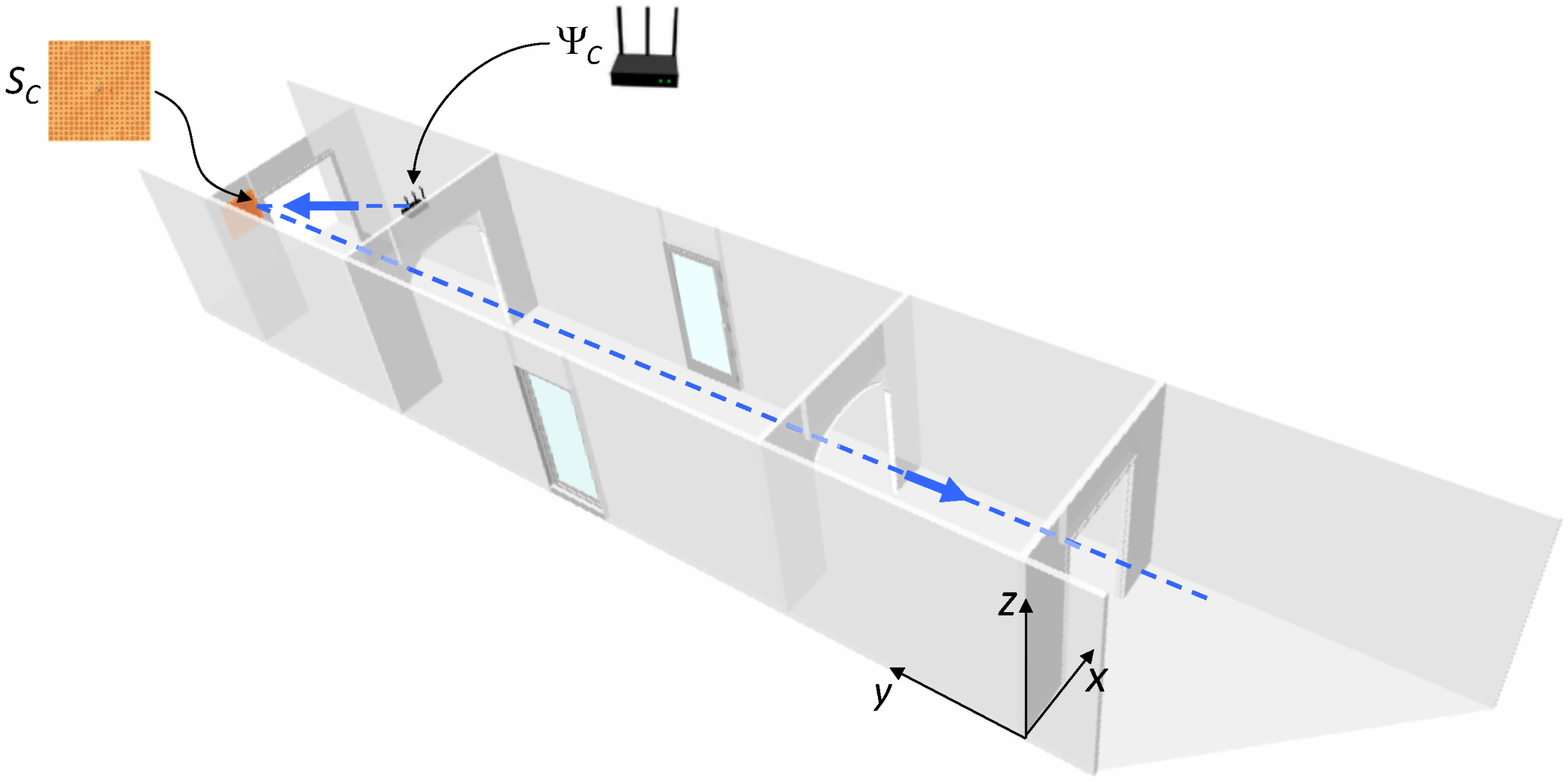}}\tabularnewline
\multicolumn{2}{c}{(\emph{a})}\tabularnewline
\multicolumn{2}{c}{}\tabularnewline
\multicolumn{2}{c}{\includegraphics[%
  width=0.60\columnwidth]{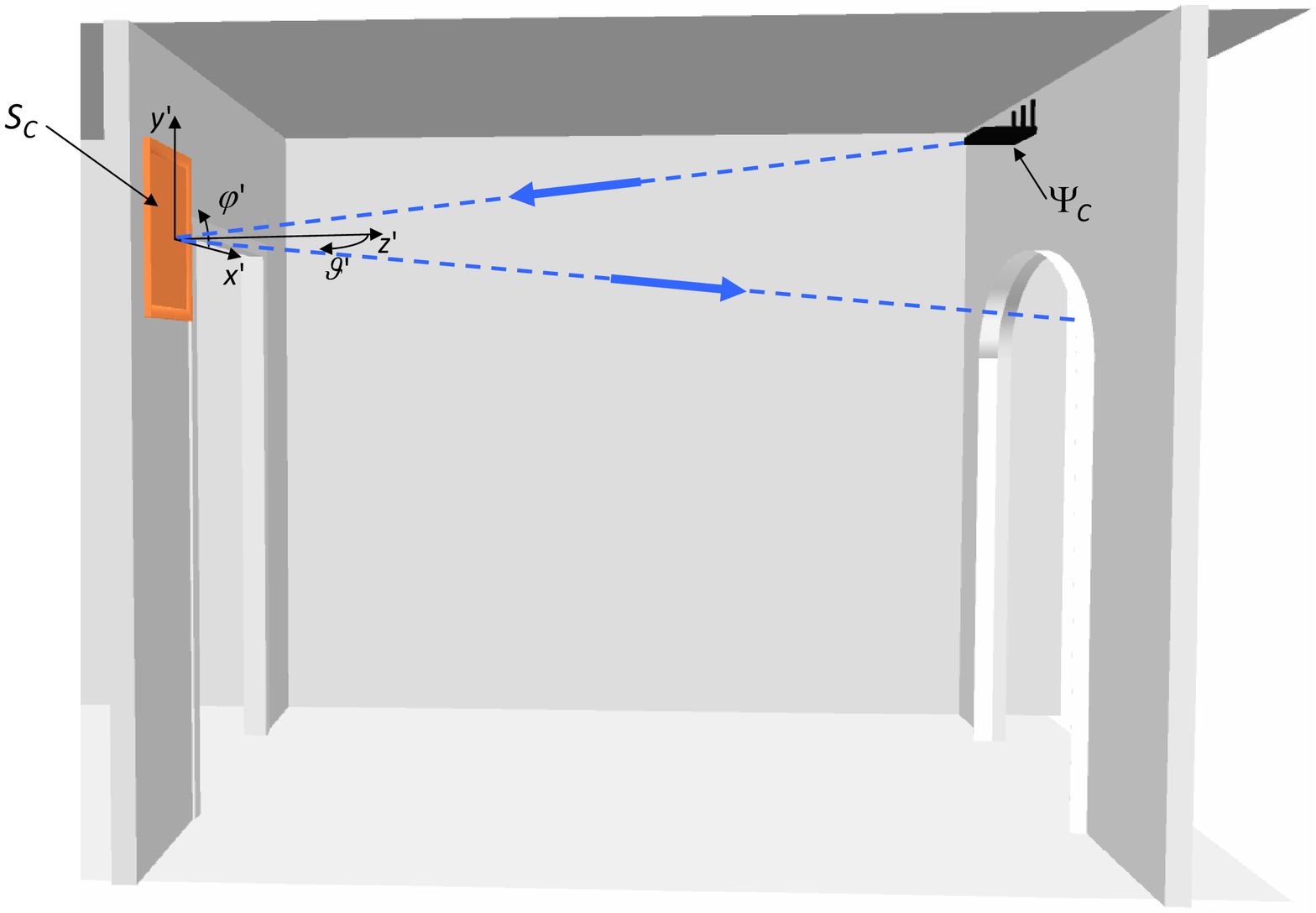}}\tabularnewline
\multicolumn{2}{c}{(\emph{b})}\tabularnewline
\end{tabular}\end{center}

\begin{center}~\vfill\end{center}

\begin{center}\textbf{Fig. 15 - A. Benoni et} \textbf{\emph{al.}}\textbf{,}
\textbf{\emph{{}``}}Towards Real-World Indoor Smart ...''\end{center}

\newpage
\begin{center}\begin{tabular}{cc}
\multicolumn{2}{c}{\includegraphics[%
  width=0.65\columnwidth]{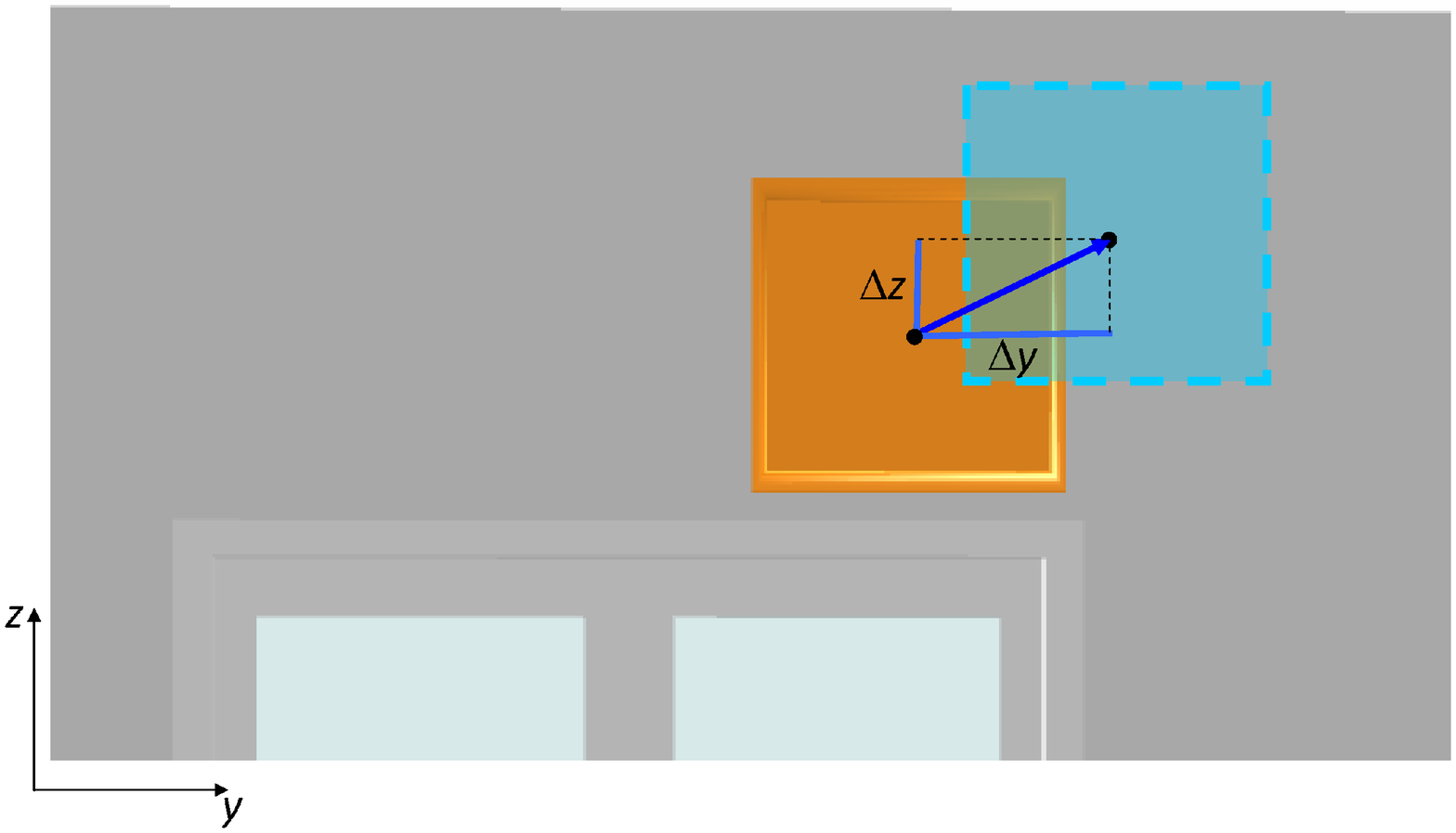}}\tabularnewline
\multicolumn{2}{c}{(\emph{a})}\tabularnewline
\multicolumn{2}{c}{}\tabularnewline
\multicolumn{2}{c}{\includegraphics[%
  width=0.45\columnwidth]{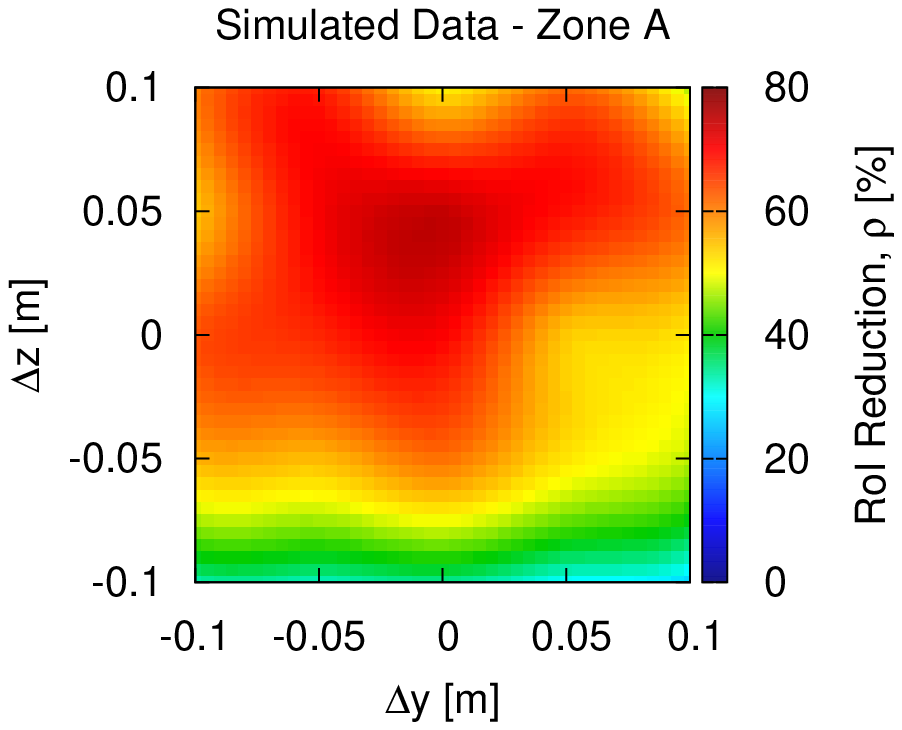}}\tabularnewline
\multicolumn{2}{c}{(\emph{b})}\tabularnewline
\multicolumn{2}{c}{}\tabularnewline
\includegraphics[%
  width=0.50\columnwidth]{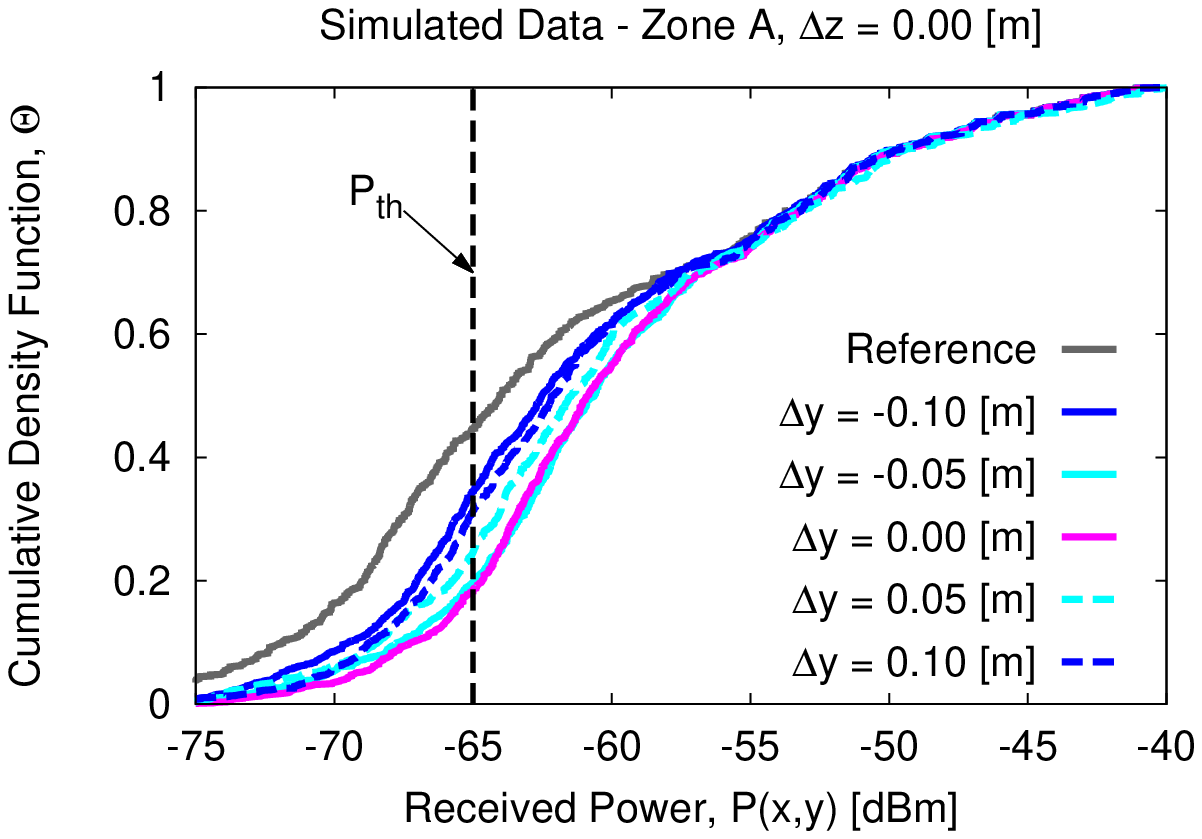}&
\includegraphics[%
  width=0.50\columnwidth]{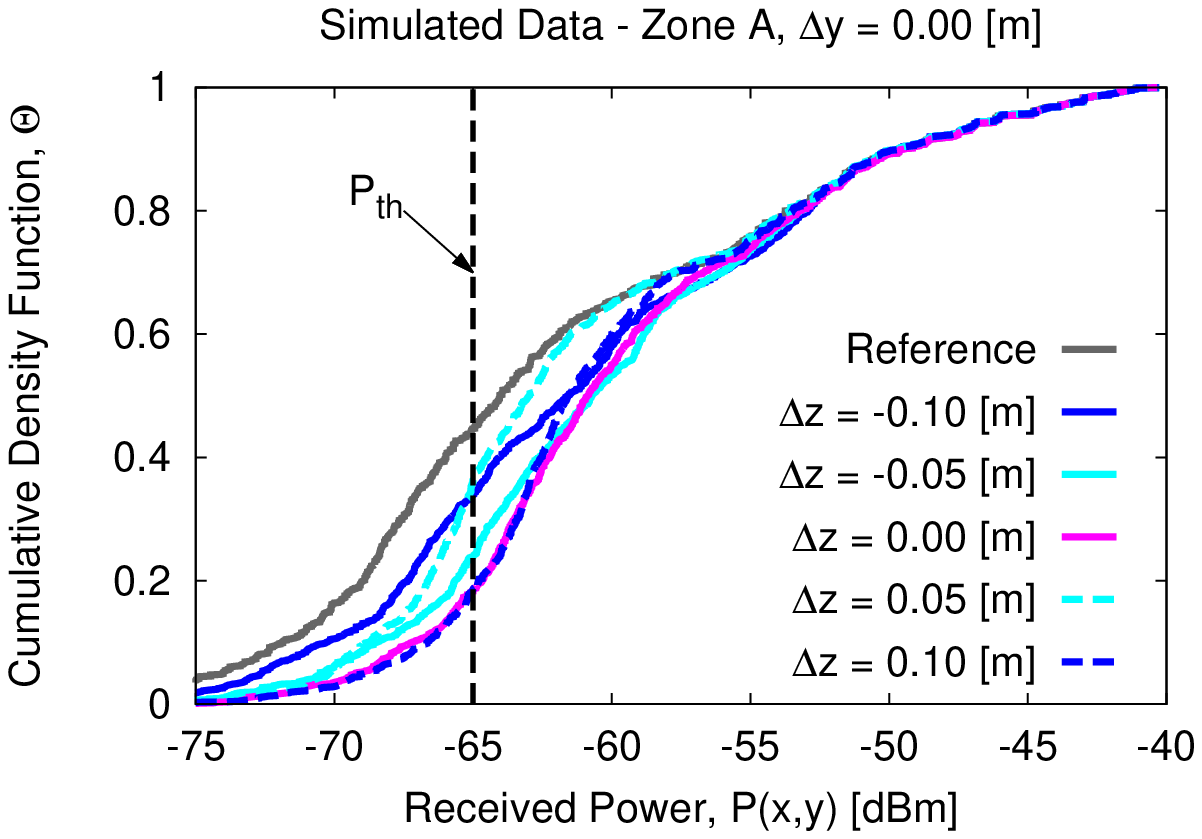}\tabularnewline
(\emph{c}) &
(\emph{d}) \tabularnewline
\end{tabular}\end{center}

\begin{center}~\vfill\end{center}

\begin{center}\textbf{Fig. 16 - A. Benoni et} \textbf{\emph{al.}}\textbf{,}
\textbf{\emph{{}``}}Towards Real-World Indoor Smart ...''\end{center}

\newpage
\begin{center}~\vfill\end{center}

\begin{center}\begin{tabular}{c}
\includegraphics[%
  width=0.90\columnwidth]{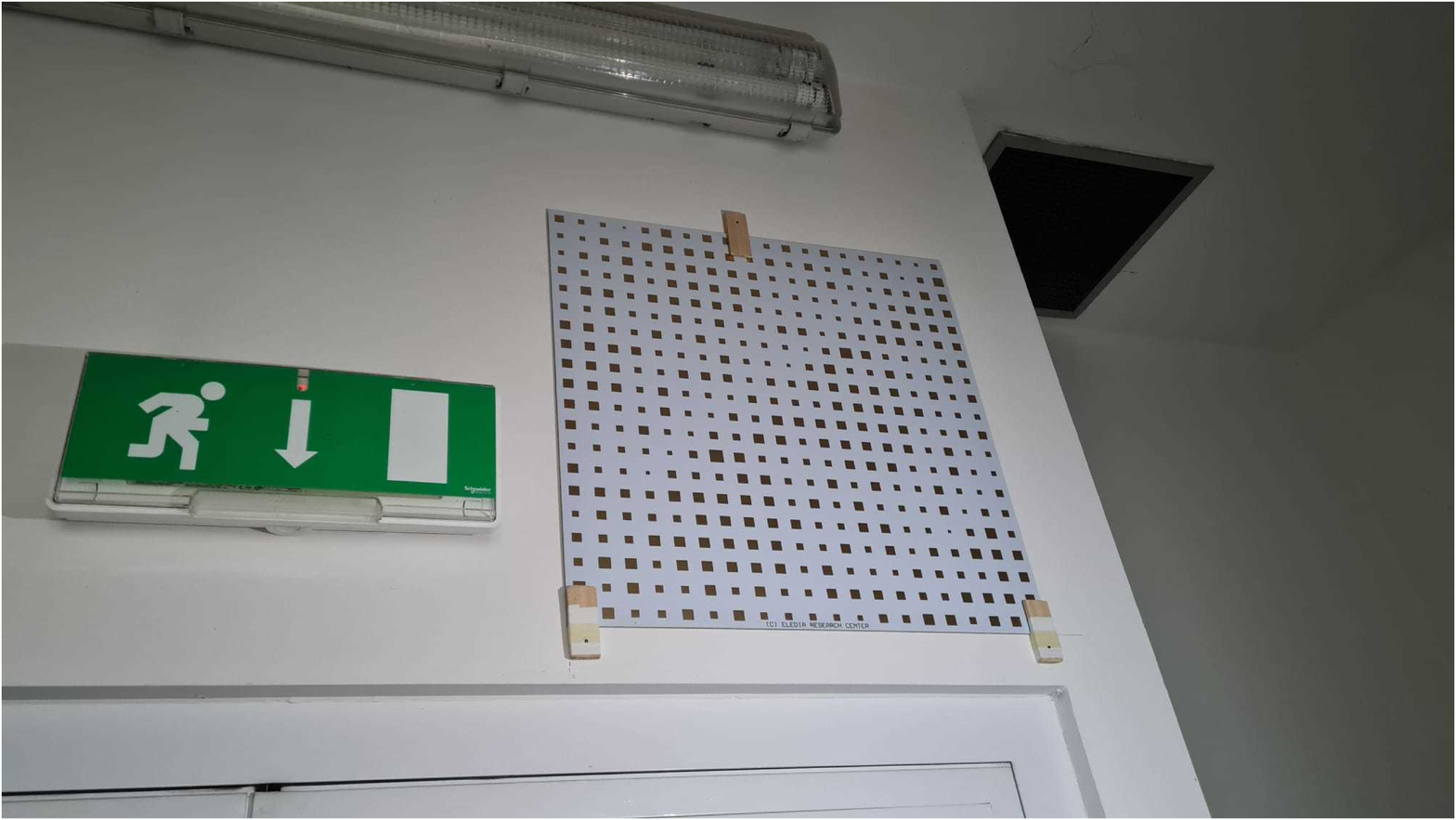}\tabularnewline
(\emph{a})\tabularnewline
\tabularnewline
\includegraphics[%
  width=0.90\columnwidth]{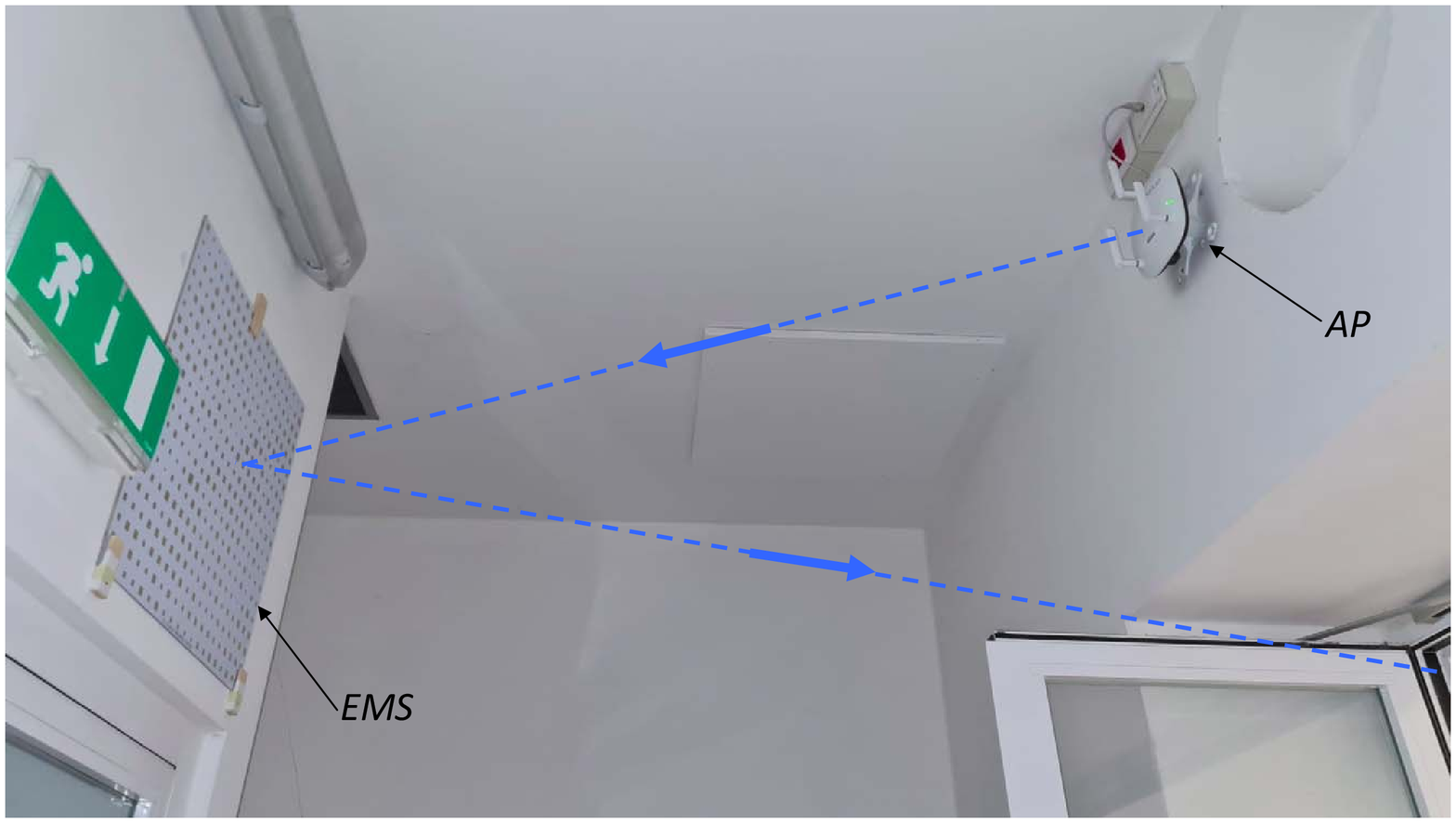}\tabularnewline
(\emph{b}) \tabularnewline
\end{tabular}\end{center}

\begin{center}~\vfill\end{center}

\begin{center}\textbf{Fig. 17 - A. Benoni et} \textbf{\emph{al.}}\textbf{,}
\textbf{\emph{{}``}}Towards Real-World Indoor Smart ...''\end{center}

\newpage
\begin{center}~\vfill\end{center}

\begin{center}\begin{tabular}{c}
\includegraphics[%
  width=0.90\columnwidth]{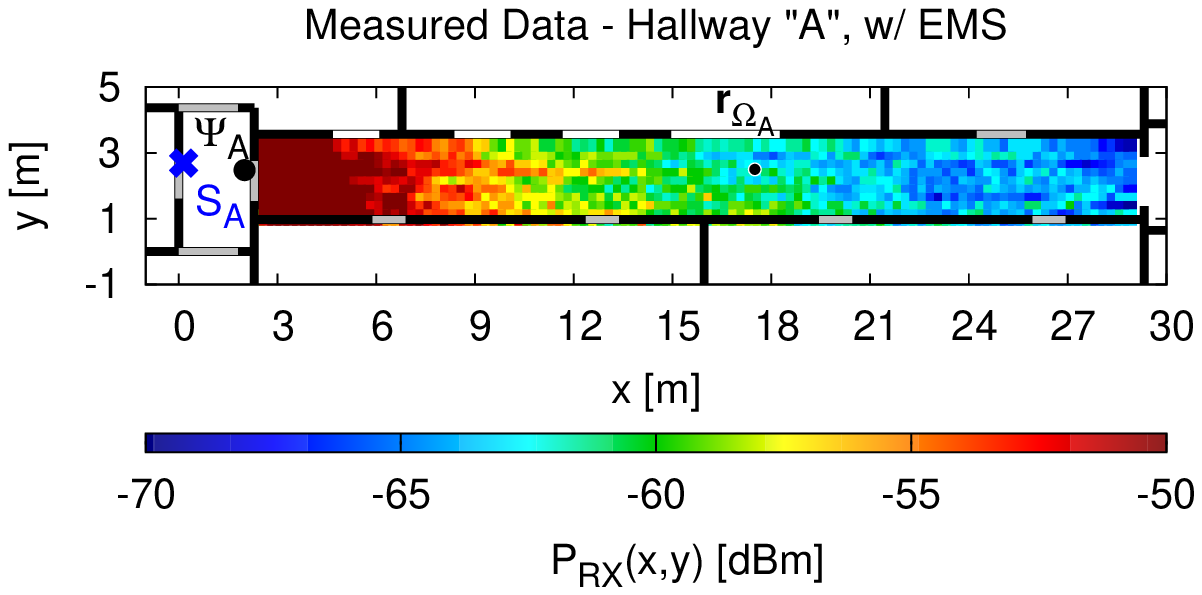}\tabularnewline
(\emph{a})\tabularnewline
\tabularnewline
\includegraphics[%
  width=0.90\columnwidth]{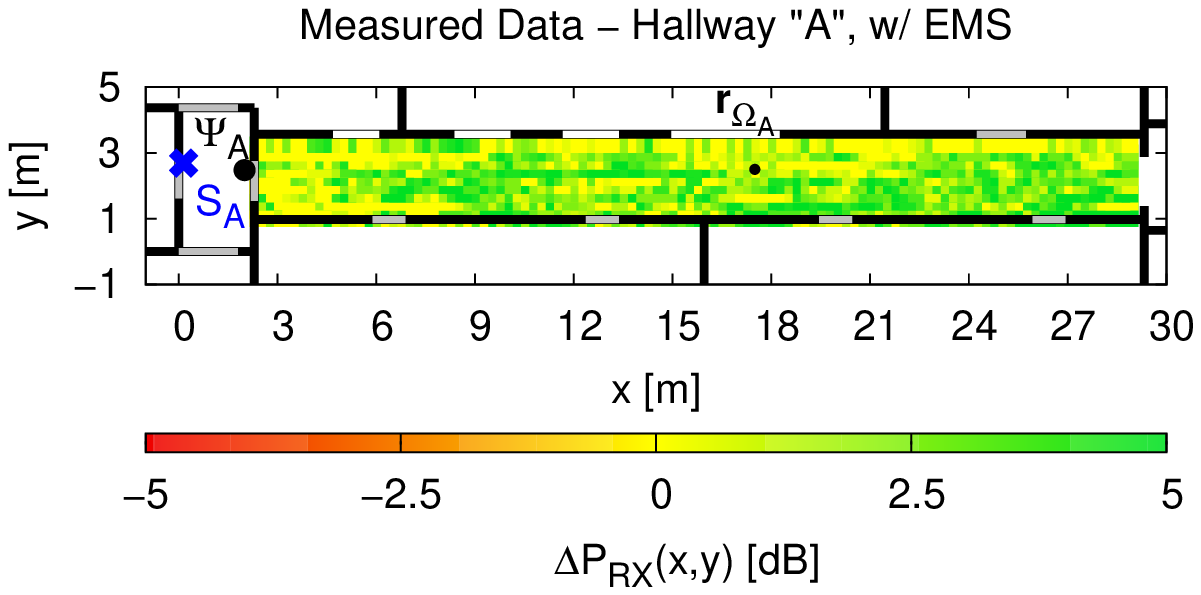}\tabularnewline
(\emph{b})\tabularnewline
\end{tabular}\end{center}

\begin{center}~\vfill\end{center}

\begin{center}\textbf{Fig. 18 - A. Benoni et} \textbf{\emph{al.}}\textbf{,}
\textbf{\emph{{}``}}Towards Real-World Indoor Smart ...''\end{center}

\newpage
\begin{center}~\vfill\end{center}

\begin{center}\includegraphics[%
  width=0.75\columnwidth]{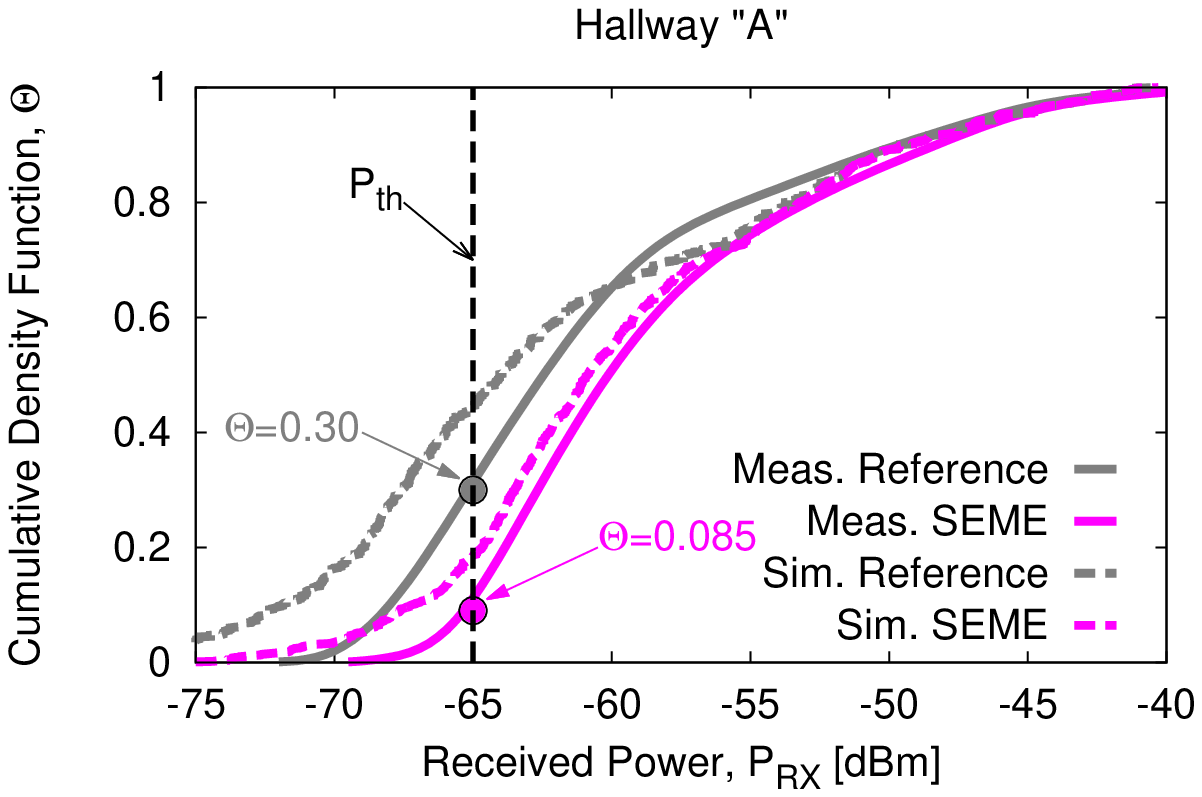}\end{center}

\begin{center}~\vfill\end{center}

\begin{center}\textbf{Fig. 19 - A. Benoni et} \textbf{\emph{al.}}\textbf{,}
\textbf{\emph{{}``}}Towards Real-World Indoor Smart ...''\end{center}

\newpage
\begin{center}~\vfill\end{center}

\begin{center}\begin{tabular}{ccc}
\begin{sideways}
\end{sideways}&
Download Throughput ($\mathcal{D}$)&
Upload Throughput ($\mathcal{U}$)\tabularnewline
\begin{sideways}
\emph{~~~~~~~~~~~~~~Reference}%
\end{sideways}&
\includegraphics[%
  width=0.45\columnwidth]{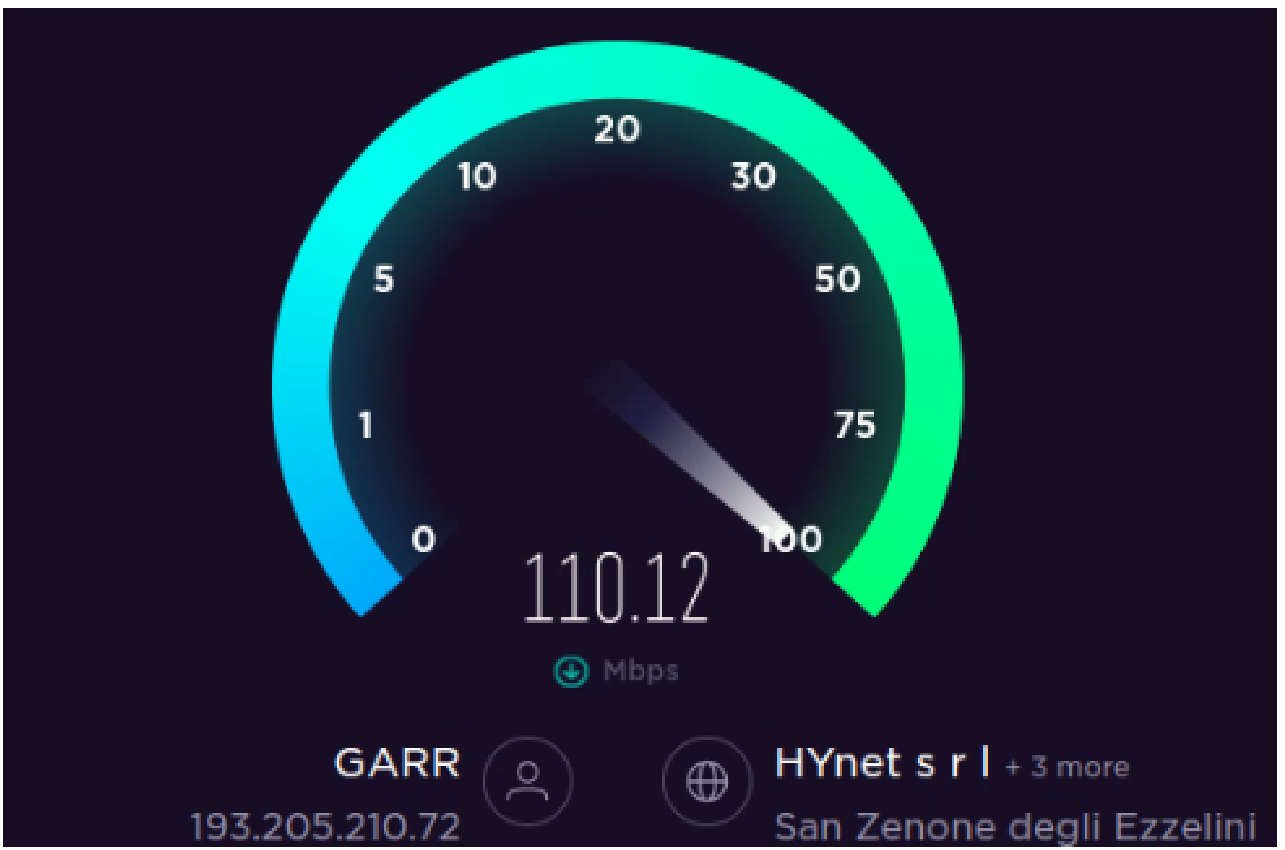}&
\includegraphics[%
  width=0.45\columnwidth]{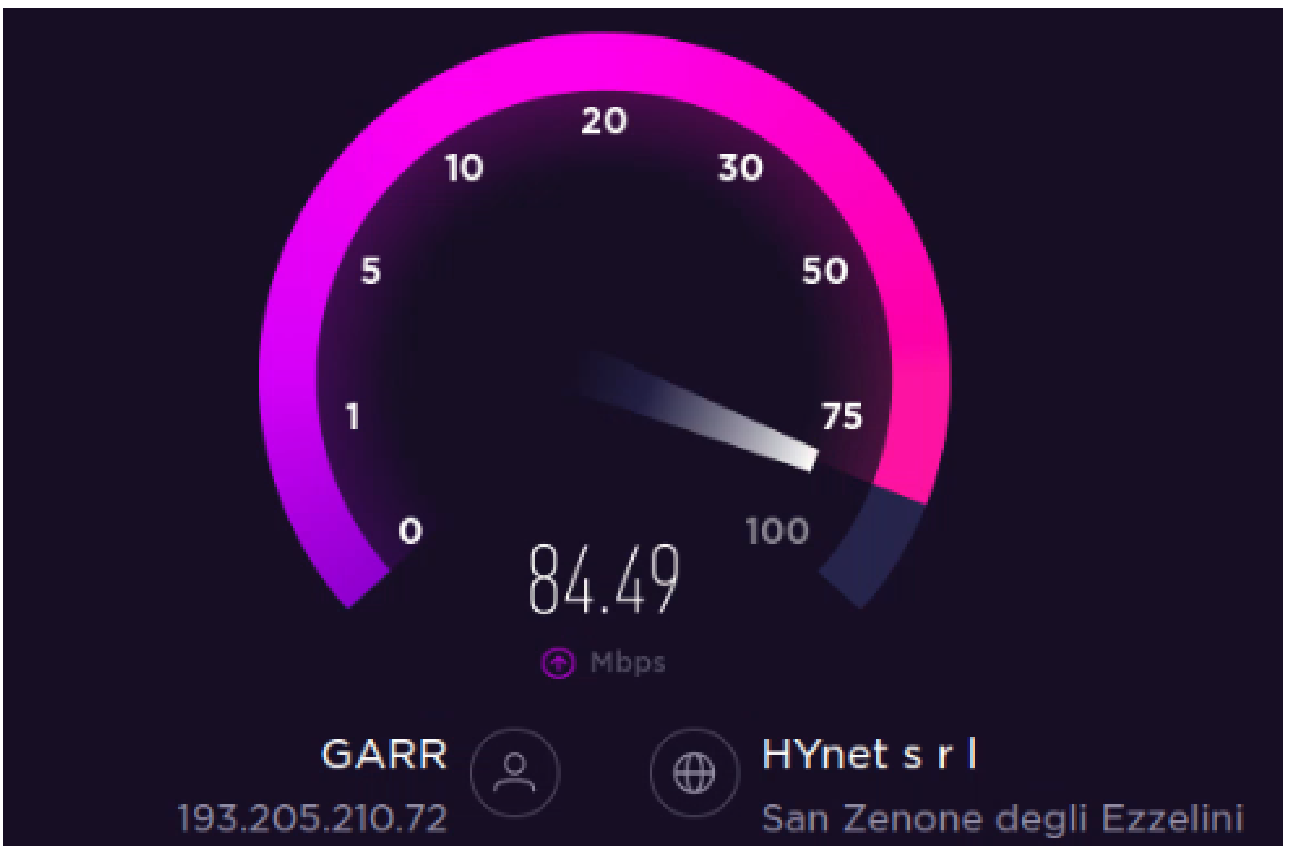}\tabularnewline
\begin{sideways}
\end{sideways}&
(\emph{a})&
(\emph{b})\tabularnewline
\begin{sideways}
\end{sideways}&
&
\tabularnewline
\begin{sideways}
\emph{~~~~~~~~~~~~~~~SEME}%
\end{sideways}&
\includegraphics[%
  width=0.45\columnwidth]{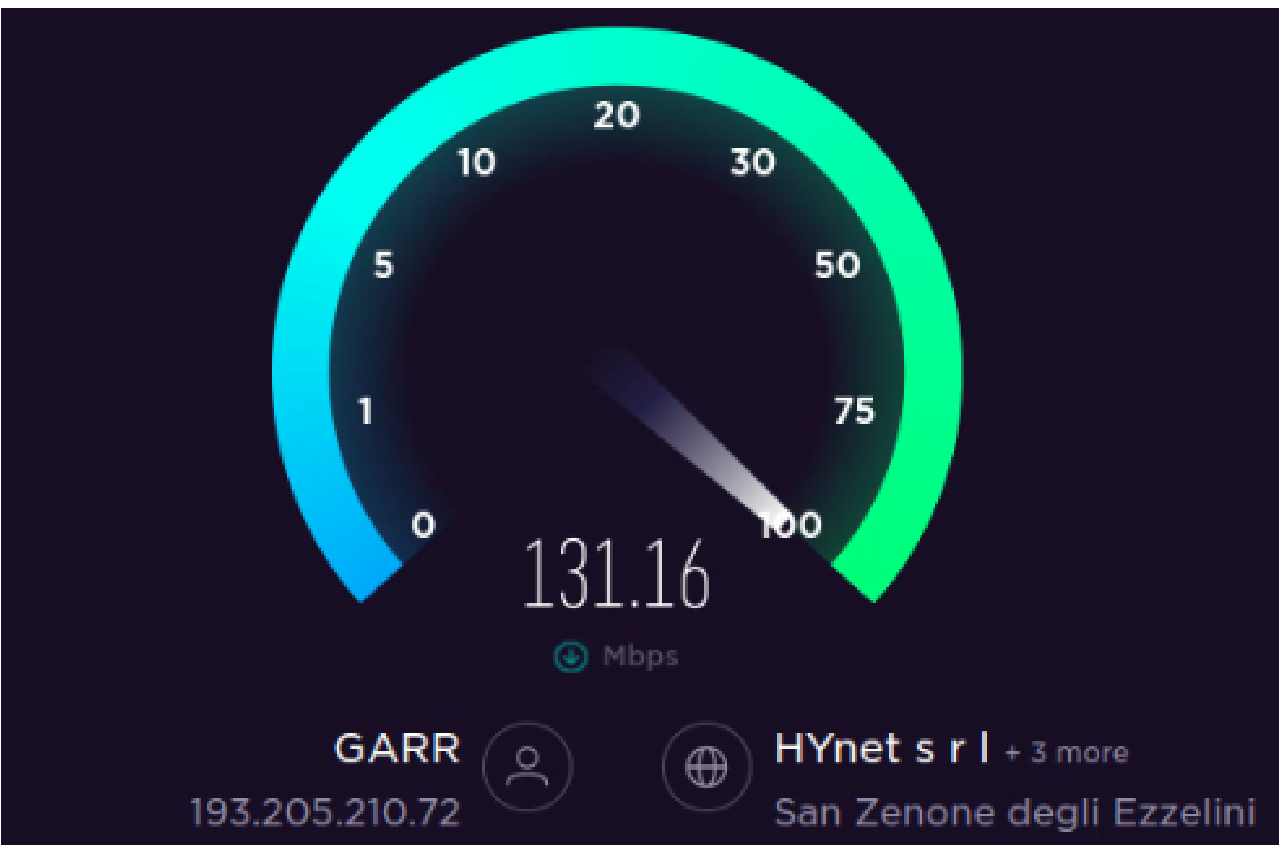}&
\includegraphics[%
  width=0.45\columnwidth]{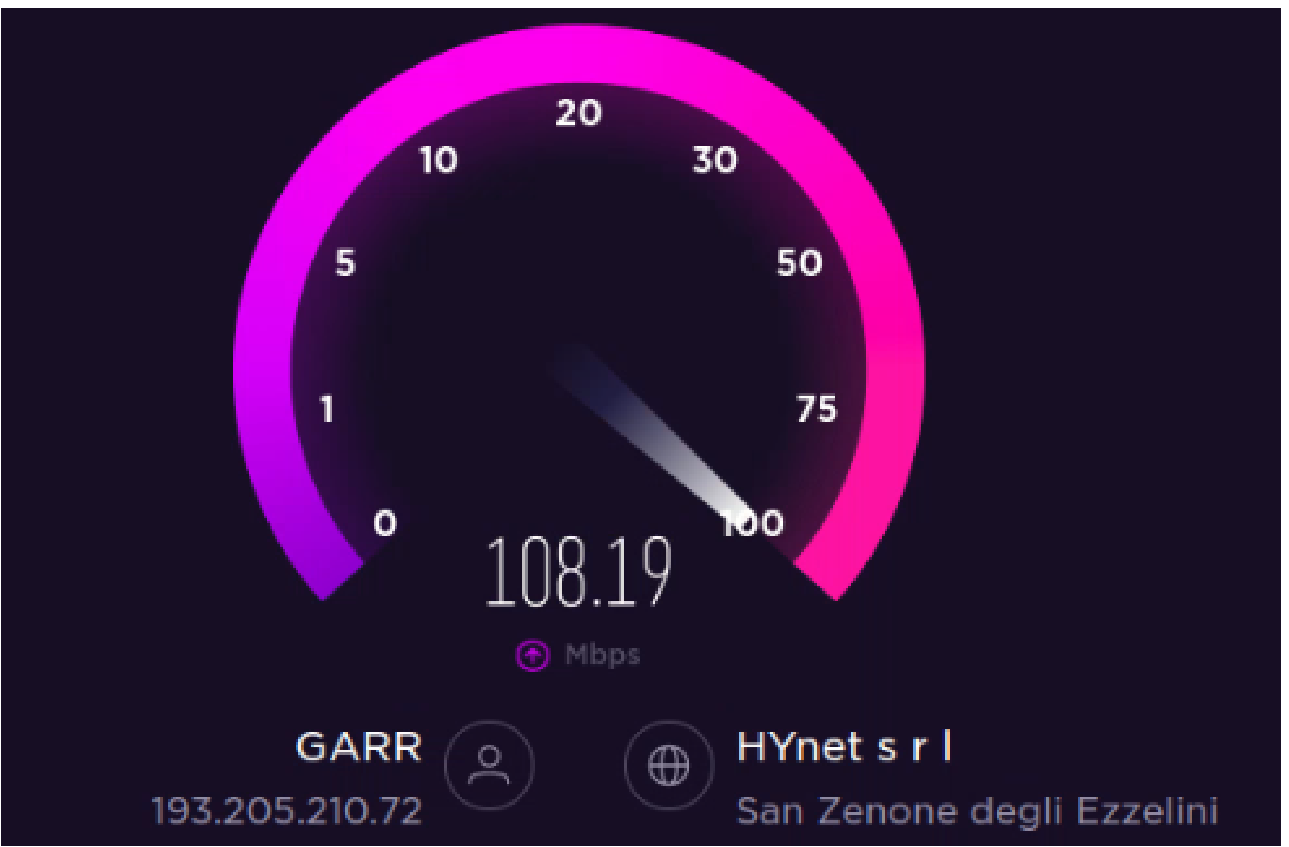}\tabularnewline
\begin{sideways}
\end{sideways}&
(\emph{c})&
(\emph{d})\tabularnewline
\end{tabular}\end{center}

\begin{center}~\vfill\end{center}

\begin{center}\textbf{Fig. 20 - A. Benoni et} \textbf{\emph{al.}}\textbf{,}
\textbf{\emph{{}``}}Towards Real-World Indoor Smart ...''\end{center}

\newpage
\begin{center}~\vfill\end{center}

\begin{center}\begin{tabular}{|c||c|c||c|c|c|c|}
\hline 
\emph{EMS} Side&
\emph{RoI} Area&
\emph{RoI} Reduction&
\textbf{$\Delta P_{RX}^{\min}$}&
\textbf{$\Delta P_{RX}^{\max}$}&
\textbf{$\Delta P_{RX}^{\mathrm{avg}}$ }&
\textbf{$\Delta P_{RX}^{\mathrm{dev}}$}\tabularnewline
$L_{A}$ {[}m{]}&
$\Lambda_{SEME}\left(\Omega_{A}\right)$ {[}$\mathrm{m}^{2}${]}&
$\rho\left(\Omega_{A}\right)$ {[}\%{]}&
{[}dB{]}&
{[}dB{]}&
{[}dB{]}&
{[}dB{]}\tabularnewline
\hline
\hline 
$0.28$&
$14.38$&
$38.84$&
$0.00$&
$5.74$&
$1.18$&
$1.02$\tabularnewline
\hline 
$0.40$&
$9.75$&
$57.84$&
$0.00$&
$8.20$&
$1.96$&
$1.67$\tabularnewline
\hline 
$0.55$&
$6.94$&
$70.00$&
$0.00$&
$9.60$&
$2.64$&
$2.36$\tabularnewline
\hline 
$0.66$&
$4.69$&
$76.22$&
$0.00$&
$11.60$&
$3.27$&
$2.95$\tabularnewline
\hline 
$0.80$&
$4.75$&
$79.46$&
$0.00$&
$13.21$&
$3.66$&
$3.46$\tabularnewline
\hline
\end{tabular}\end{center}

\begin{center}~\vfill\end{center}

\begin{center}\textbf{Tab. I - A. Benoni et} \textbf{\emph{al.}}\textbf{,}
\textbf{\emph{{}``}}Towards Real-World Indoor Smart ...''\end{center}

\newpage
\begin{center}\begin{sideways}
\begin{tabular}{|c||c|c|c|c|c||c|c|c|c|}
\hline 
\emph{Hallway}&
\emph{Area }&
\emph{RoI} Area&
\emph{EMS} Side&
\emph{RoI} Area&
\emph{RoI} Reduction&
\textbf{$\Delta P_{RX}^{\min}$ }&
\textbf{$\Delta P_{RX}^{\max}$ }&
\textbf{$\Delta P_{RX}^{\mathrm{avg}}$ }&
\textbf{$\Delta P_{RX}^{\mathrm{dev}}$} \tabularnewline
\hline
\hline 
$\gamma$&
{[}$\mathrm{m}^{2}${]}&
$\Lambda_{Ref}\left(\Omega\right)$ {[}$\mathrm{m}^{2}${]}&
$L${[}$\mathrm{m}${]}&
$\Lambda_{SEME}\left(\Omega\right)$ {[}$\mathrm{m}^{2}${]}&
$\rho\left(\Omega\right)$ {[}\%{]}&
{[}\emph{dB}{]}&
{[}\emph{dB}{]}&
{[}\emph{dB}{]}&
{[}\emph{dB}{]}\tabularnewline
\hline
\hline 
\emph{B}&
$67.5$&
$21.88$&
$0.55$&
$6.06$&
$72.29$&
$0.00$&
$8.21$&
$2.52$&
$1.78$\tabularnewline
\hline 
\emph{C}&
$60.0$&
$0.8$&
$0.32$&
$0.00$&
$100$&
$0.00$&
$3.87$&
$0.84$&
$0.51$\tabularnewline
\hline 
\emph{D}&
$40.5$&
-&
-&
-&
-&
-&
-&
-&
-\tabularnewline
\hline 
\emph{E}&
$40.5$&
-&
-&
-&
-&
-&
-&
-&
-\tabularnewline
\hline
\end{tabular}
\end{sideways}\end{center}

\begin{center}\textbf{Tab. II - A. Benoni et} \textbf{\emph{al.}}\textbf{,}
\textbf{\emph{{}``}}Towards Real-World Indoor Smart ...''\end{center}

\newpage
\begin{center}~\vfill\end{center}

\begin{center}\begin{tabular}{|c||c|c|c|c||c|c|c|c||c|}
\hline 
&
\multicolumn{4}{c||}{\emph{Reference}}&
\multicolumn{4}{c||}{\emph{SEME}}&
\tabularnewline
\cline{2-5} \cline{6-9} 
&
$\min$&
$\max$&
avg&
std&
$\min$&
$\max$&
avg&
std&
$\Xi$\tabularnewline
\hline
\hline 
$\mathcal{D}$ {[}Mbps{]}&
$74.94$&
$110.12$&
$95.79$&
$8.87$&
$109.60$&
$131.16$&
$127.98$&
$5.91$&
$33.6$ \%\tabularnewline
\hline 
$\mathcal{L}_{\mathcal{D}}$ {[}msec{]}&
$18.00$&
$189.00$&
$122.87$&
$50.92$&
$16.00$&
$73.00$&
$52.13$&
$15.99$&
$-57.6$ \%\tabularnewline
\hline
\hline 
$\mathcal{U}$ {[}Mbps{]}&
$77.14$&
$84.49$&
$81.57$&
$2.62$&
$80.54$&
$108.19$&
$97.18$&
$9.79$&
$19.1$ \%\tabularnewline
\hline 
$\mathcal{L}_{\mathcal{U}}$ {[}msec{]}&
$24.00$&
$116.00$&
$43.33$&
$23.58$&
$11.00$&
$25.00$&
$20.40$&
$4.00$&
$-52.9$ \%\tabularnewline
\hline
\end{tabular}\end{center}

\begin{center}~\vfill\end{center}

\begin{center}\textbf{Tab. III - A. Benoni et} \textbf{\emph{al.}}\textbf{,}
\textbf{\emph{{}``}}Towards Real-World Indoor Smart ...''\end{center}

\newpage
\begin{center}~\vfill\end{center}

\begin{center}\begin{tabular}{|c||c|c|c|}
\hline 
&
\multicolumn{3}{c|}{$\Delta$\emph{Cost} (\emph{w.r.t. Reference})}\tabularnewline
\cline{2-4} 
&
\emph{Reference} (\emph{AP})&
\emph{SEME} (\emph{AP} \& \emph{SP-EMS})&
\emph{Standard} (2 \emph{AP}s)\tabularnewline
\hline
\hline 
\emph{Device Purchase} ($\mathcal{C}_{a}$) {[}\${]}&
$0$&
$100$&
$700$\tabularnewline
\hline 
\emph{Device Installation} ($\mathcal{C}_{c}$) {[}\${]}&
$0$&
$5$&
$300$\tabularnewline
\hline 
\emph{Energy Cost} ($\mathcal{C}_{o}$) {[}\$/year{]}&
$0$&
$0$&
$50$\tabularnewline
\hline
\emph{Device Removal} ($\mathcal{C}_{d}$) {[}\${]}&
$0$&
$0$&
$100$\tabularnewline
\hline
\hline 
\emph{TCO~}($\mathbb{C}$)&
$0$ {[}\${]} + $0$ {[}\$/year{]}&
$105$ {[}\${]} + $0$ {[}\$/year{]}&
$1100$ {[}\${]} + $50$ {[}\$/year{]}\tabularnewline
\hline
\hline 
\emph{5-Years TCO} {[}\${]}&
$0$&
$105$&
$1350$\tabularnewline
\hline
\end{tabular}\end{center}

\begin{center}~\vfill\end{center}

\begin{center}\textbf{Tab. IV - A. Benoni et} \textbf{\emph{al.}}\textbf{,}
\textbf{\emph{{}``}}Towards Real-World Indoor Smart ...''\end{center}
\end{document}